# An Unobtrusive and Lightweight Ear-worn System for Continuous Epileptic Seizure Detection

ABDUL AZIZ, College Information and Computer Sciences, University of Massachusetts Amherst, USA
NHAT PHAM, School of Computer Science and Informatics, Cardiff University, UK
NEEL VORA, Department of Computer Science & Engineering, University of Texas at Arlington, USA
CODY REYNOLDS, Department of Computer Science & Engineering, University of Texas at Arlington, USA
JAIME LEHNEN, Department of Neurology, The University of Texas Southwestern Medical Center, USA
POOJA VENKATESH, Department of Neurology, The University of Texas Southwestern Medical Center, USA
ZHUORAN YAO, Department of Neurology, The University of Texas Southwestern Medical Center, USA
JAY HARVEY, Department of Neurology, The University of Texas Southwestern Medical Center, USA
TAM VU, Earable Inc., USA
KAN DING, Department of Neurology, The University of Texas Southwestern Medical Center, USA
PHUC NGUYEN*, College Information and Computer Sciences, University of Massachusetts Amherst, USA

Epilepsy is one of the most common neurological diseases globally (around 50M people globally). Fortunately, up to 70% of people with epilepsy could live seizure-free if properly diagnosed and treated, and a reliable technique to monitor the onset of seizures could improve the quality of life of patients who are constantly facing the fear of random seizure attacks. The current gold standard, video-EEG (v-EEG), involves attaching over 20 electrodes to the scalp, is costly, requires hospitalization, trained professionals, and is uncomfortable for patients. To address this gap, we developed *EarSD*, a lightweight and unobtrusive ear-worn system to detect seizure onsets by measuring physiological signals behind the ears. This system can be integrated into earphones, headphones, or hearing aids, providing a convenient solution for continuous monitoring. *EarSD* is an integrated custom-built *sensing-computing-communication* ear-worn platform to capture seizure signals, remove the noises caused by motion artifacts and environmental impacts, and stream the collected data wirelessly to the computer/mobile phone nearby. *EarSD*'s ML algorithm, running on a server, identifies seizure-associated signatures and detects onset events. We evaluated the proposed system in both in-lab and in-hospital experiments at the University of Texas Southwestern Medical Center with epileptic seizure patients, confirming its usability and practicality.

## 1 INTRODUCTION

**Motivations.** Epileptic seizures are one of the most prevalent neurological disorders affecting approximately 50 million people worldwide, with an estimated 5 million new cases diagnosed every year [1]. Seizures can occur suddenly and unpredictably, leading to severe accidents or even death [2]. Fortunately, up to 70% of people with epilepsy could live seizure-free if properly diagnosed and treated. A reliable and effective prediction technique to

*Corresponding author.

Authors' addresses: Abdul Aziz, College Information and Computer Sciences, University of Massachusetts Amherst, Amherst, Massachusetts, USA, abdulaziz@umass.edu; Nhat Pham, School of Computer Science and Informatics, Cardiff University, Cardiff, UK, phamn@cardiff.ac.uk; Neel Vora, Department of Computer Science & Engineering, University of Texas at Arlington, Arlington, Texas, USA, nxv8988@mavs.uta.edu; Cody Reynolds, Department of Computer Science & Engineering, University of Texas at Arlington, Arlington, Texas, USA, cody.reynolds2@mavs.uta.edu; Jaime Lehnen, Department of Neurology, The University of Texas Southwestern Medical Center, Dallas, Texas, USA, jaime.lehnen@utsouthwestern.edu; Pooja Venkatesh, Department of Neurology, The University of Texas Southwestern Medical Center, Dallas, Texas, USA, pooja.venkatesh@utsouthwestern.edu; Zhuoran Yao, Department of Neurology, The University of Texas Southwestern Medical Center, Dallas, Texas, USA, zhuoran.yao@utsouthwestern.edu; Jay Harvey, Department of Neurology, The University of Texas Southwestern Medical Center, Dallas, Texas, USA, jay.harvey@utsouthwestern.edu; Tam Vu, Earable Inc., Boulder, Colorado, USA, tam@earable.ai; Kan Ding, Department of Neurology, The University of Texas Southwestern Medical Center, Dallas, Texas, USA, kan.ding@utsouthwestern.edu; Phuc Nguyen, College Information and Computer Sciences, University of Massachusetts Amherst, Amherst, Massachusetts, USA, vp.nguyen@cs.umass.edu.

anticipate the onset of seizures could improve the quality of life of patients who are constantly facing the fear of random seizure attacks. An epileptic seizure is a type of seizure that results in a temporary loss of control accompanied by convulsions, unconsciousness, or both. Epileptic seizure generates sudden abnormal electrical discharges at different locations in the human brain. These signals can be captured using a sensitive bio-signal monitoring system.

**Prior Works and Their Limitations.** In hospitals, video Electroencephalogram (video EEG/vEEG) is considered the gold standard for diagnosing and detecting the onset of epileptic seizures by recording EEG signals from the patient's head. In particular, patients are required to spend several days (up to a week) in a hospital, mostly at Epilepsy Monitoring Units in the U.S., for a vEEG test. During the test, they wear a headset with over 20 wired electrodes to monitor electrical activity in the brain. The patients are under constant video surveillance so doctors can review the recordings to identify events that might have triggered a seizure. The collected data are used to diagnose seizures. There has been continuous research to develop automated seizure detection tools to improve the reliability of EEG-based seizure monitoring over the last 50 years [3]. Most efforts have been devoted to developing robust seizure detection algorithms using signal processing, feature extraction, and machine learning techniques[4, 5] based on the collected vEEG data. While these works have demonstrated seizure classification accuracy of over 90%, vEEG setup is uncomfortable for users and needs to be set up and operated by trained technicians. Moreover, the study is costly, making long-term data collection unfeasible. More importantly, due to the rareness of the diseases, many patients went home with zero seizure events detected during hospitalization.

Indeed, seizure data collected at users' homes before and after hospitalization is critical for monitoring disease progression and treatment. However, there is currently no robust and practical solution for analyzing their neurological condition beyond their short hospital stay. When patients are at home, current methods predominantly rely on self-reporting through seizure diaries. Studies have shown, however, that seizure records collected in this manner are accurate for only about 50% of patients [6]. The absence of reliable, wearable devices that patients can use in their daily lives and under any environments or situations means that many seizures go unrecorded, and the contextual information that could have triggered the event remains incomplete. Doctors have to rely on this incomplete and subjective information which impacts the treatment method. Therefore, there is a dire need for a portable, wearable system that can enable continuous, long-term monitoring to provide consistent and reliable data to doctors for better treatment.

Leveraging advancements in the development of miniaturized sensors and electronics, improved wireless data transmission techniques, and rechargeable batteries, ongoing research has focused on developing mobile, home monitoring solutions for seizure detection. A new wrist-worn device, namely Empatica Embrace2 [7], has been proven to detect a few types of seizure reliably. However, it is only robust with very limited types of seizure [8], mostly tonic-clonic, as they only capture signals around the wrist areas. Neuronaute EEG System/IceCap EEG System [9] is a headcap-based system for seizure monitoring, but the device is cumbersome and not socially acceptable to wear in public. Other headband devices that can capture brain signals, such as Frenz Brainband [10], Emotiv [11], NeuroSky MindWave [12], BrainLink Pro [13], Muse [14], Versus [15], Neuroon Open [16], Naptime [17], etc., are primarily designed for meditation, sleep improvement, and wellness tracking [18]. These are head-band form factors, which are different from our wearable settings. In addition, they use EEG signals to provide feedback on brain states to help users maintain focus during meditation or improve sleep quality. They are not designed, optimized, and clinically validated for seizure detection. Our goal is to integrate *EarSD* into existing earable systems that are already socially acceptable.

**Proposed Approach.** In this paper, we explore a novel and robust sensing system integrated into one of the most well-accepted wearable form factors – *the everyday earbuds*– for epileptic seizure detection, as illustrated in Fig. 1. Our proposed device, namely *EarSD*, collects physiological signals of EEG, EMG, and EOG from behind the ear and fetches them into machine learning models to accurately and rapidly detect seizure onsets. Unlink existing

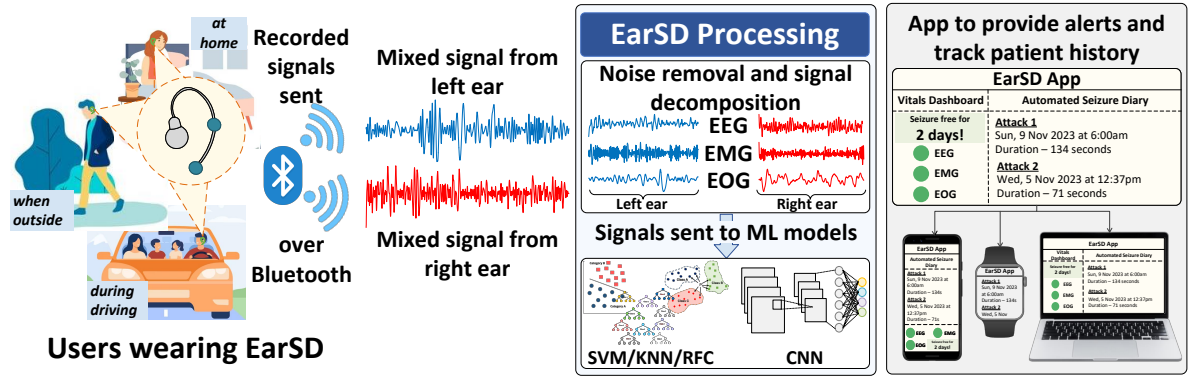

Fig. 1. The vision of *EarSD*, a socially acceptable wearable that supports real-time epileptic seizure detection. *EarSD* will be worn by epileptic seizure patients pre- and post-hospitalization. *EarSD* can be used as a standalone device or combined with video EEG to allow caregivers to design an effective treatment plan.

EEG headband devices that focus on the forehead area; this work validates the feasibility of developing a fully integrated ear-worn system for seizure monitoring. Thanks to the social acceptance of earbuds/earphones, *EarSD* could be worn in all environments, making it an ideal solution for continuous patient monitoring. *EarSD* has the potential to revolutionize the field of epilepsy management and significantly improve the quality of life for those affected by the condition. First, timely detection of epilepsy is critical for prompt intervention and mitigating potential risks and complications. Second, *EarSD* can enable real-time detection for early warnings and facilitate swift medical care. Third, the inconspicuous nature of an ear-worn wearable enhances patient compliance and encourages long-term usage, leading to more comprehensive data collection. Fourth, such unobtrusive devices will empower individuals to live their lives more freely, with the confidence of timely seizure detection and improved seizure management. Last but not least, by collecting long-term data using *EarSD*, we can contribute to a deeper understanding of epilepsy and facilitate personalized treatment plans.

Besides epileptic seizure monitoring, if successful, *EarSD* system can be considered as a reference design for many other brain disorders monitoring systems, including. Neuromuscular Diseases [19], Autism and Neurodevelopment [20], dementia [21], pain [22], movement disorders [23], and others.

**Challenges.** However, realizing *EarSD* is difficult due to the following challenges.

- The relationship between the signals from around the ear and epileptic seizure onset has not been *thoroughly* and *clinically* analyzed in the literature. Note that no electrodes have been placed exactly around the ear locations in standard vEEG settings.
- Developing a sensitive *wearable* system capable of accurately capturing head-based signals in a compact design is challenging. To be specific, a critical obstacle is devising a resilient method that can effectively eliminate the influence of human-generated disturbances on the monitored signals. For instance, while brain (EEG), muscle (EMG), and eye (EOG) signals typically range from microvolts ($\mu V$) to a few millivolts ($mV$), bodily movements like head motion, walking, or talking can significantly overshadow the sensor data, causing noise levels to spike up to several volts.
- Performing real-time data acquisition, signal processing, and computing on resource-constrained wearable devices is a demanding task limiting their usefulness for timely and effective seizure management. Efficient frameworks are needed to support real-time data acquisition, signal analysis, and inference while still operating within the limitations of the hardware.

- Unlike common wearable devices designed for healthy individuals, *EarSD* is designed as a medical diagnosis tool to augment and complement the hospital-based vEEG recordings so that doctors can have access to reliable recordings even when patients are not admitted to the hospital. Its accuracy, efficiency, and robustness must be evaluated in clinical settings. Developing an end-to-end research prototype, which utilizes cost-of-the-shelf hardware and software front ends, requires a comprehensive analysis and thorough engineering efforts and skills in order to approach clinical settings accurately, which is currently only obtained by tens of thousands of dollars system (EEG).

**Contributions.** This project addresses the aforementioned challenges and aims to make fundamental contributions to low-power, low-cost, unobtrusive, high-fidelity, and robust ear-based sensing systems for physiological signal monitoring. We take a holistic approach from form factor fabrication, sensing circuit design, and implementation to algorithm development to build and deploy the first ear-based epileptic seizure systems in clinical settings. We first design a sophisticated hardware and firmware pipeline to reduce the noise and then extract the mixed physiological signals collected around the ear into EEG, EMG, and EOG. We then explored multiple signal separation techniques, including ICA, PCA, EMD, and NNMF, and found that the NNMF technique is the most suitable approach. We evaluate the proposed solution on epileptic patients in a hospital to confirm the approach's feasibility, usability, and practicality. We approach this clinical accuracy (up to 97.9% accuracy) using only two electrodes behind the ear instead of the hospital vEEG [24] which generally have more than 20 electrodes placed on the scalp.

To summarize, the main contributions of this paper are:

- We develop a high-fidelity, noise-resilient, and socially acceptable EEG ear-based physiological monitoring method by creating an ear-worn system that can be safely worn behind the ear, allowing patients to wear it continuously for effective long-term EEG monitoring.
- We develop signal processing and decomposition techniques to capture physiological signals associated with epileptic seizures from data collected around the ear.
- We analyze the performance of multiple machine learning algorithms on detecting seizure onset events based on the data collected from the earable devices, confirming the feasibility, robustness, and practicality of the proposed ear-based platform.
- We conduct experiments on real epileptic seizure patients in the Epilepsy Monitoring Unit at the University of Texas Southwestern Medical Center in Dallas, Texas, U.S. The preliminary results confirm that *EarSD* is able to detect seizure with up to 97.9% accuracy on 32 patients.
- We conduct a user study on 32 patients and 9 medical doctors and caregivers. Most users found the system to be socially acceptable and easy to use, and doctors have also verified the reliability of our device.

## 2 BACKGROUND AND RELATED WORK

In this section, we present the current practice of epileptic seizure monitoring in clinical and off-site settings.

### 2.1 Clinical-based Studies

Patients diagnosed with seizures are admitted into the Epilepsy Monitoring Unit in the hospitals, where they are monitored continuously 24/7 for up to a week using the video-EEG system. During the test, the patients wear EEG head caps containing electrodes connected to the patient's scalp. A standard setup includes between 21 to 32 electrodes positioned at specific locations across the scalp following the International 10-20 system as illustrated in Fig. 2. The electrodes are connected to an EEG reader which amplifies these signals, records the brain's electrical activity, and displays them on a screen as a series of waves or patterns. This setup is often supported by video monitoring allowing the medical team to correlate the recorded brain activity with observable physical or behavioral changes, aiding in the diagnosis and identification of specific seizure types.

At the end of the test, eplieptologists interpret these recordings to diagnose conditions like epilepsy, tumors, or sleep disorders and formulate a treatment plan or further diagnostic investigations. To simplify this tedious task and reduce patient expenses, epileptic seizure detection has been an active research area since the early 1970s [3, 25]. Over the last few decades, there has been significant advancement in the field of automated epilepsy detection primarily using data from vEEG systems in hospitals.

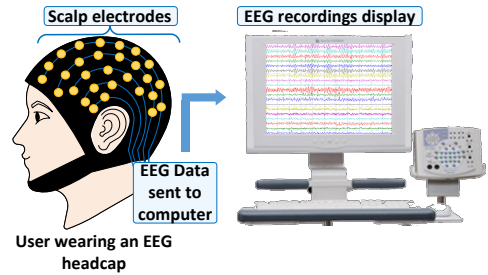

Fig. 2. Hospital-based video-EEG (vEEG) setup. Scalp electrodes capture EEG signals which are then recorded and displayed in the EEG machine for analysis by doctors.

*Improving the accuracy of physiological-based seizure detection algorithms.* Persyst's [26] algorithms for seizure detection have been widely used as a suggestion tool to medical doctors and caregivers, while it is known to have a high false positive rate [27], posing challenges for adoption in practical applications. Multiple research groups proposed various deep learning techniques for epileptic seizure detection leveraging collected video-EEG data [28]. Asif et al. [29] employed a deep learning framework, utilizing an ensemble architecture, to learn multi-spectral feature embeddings for cross-patient seizure type classification and classify seizures with 94% However, a major challenge of deep learning models is the limited availability of clinical data. Most works such as [30–33], rely on the TUSZ open seizure dataset [34].

The analysis of EEG signals is inherently complex due to the presence of noise requiring extensive preprocessing to remove unwanted artifacts. Joshi et al. [35] applied a Butterworth bandpass filter to preprocess the CHB-MIT dataset [36], another public seizure-based dataset. They performed preprocessing in both the time and frequency domains, segmenting the data into seizure and non-seizure images and processing the resultant dataset through a CNN to achieve an accuracy of 98.21%. Madhavan et al. [37] proposed an automated classification method using synchro squeezing transform (SST) and deep CNN. They transformed the one-dimensional EEG signals into two-dimensional time-frequency matrices using Fourier SST (FSST) and wavelet SST (WSST) techniques. The processed signals were then fed into a two-dimensional (2D) CNN, resulting in an accuracy of 99.94% when classifying EEG signals into focal and non-focal events. This highlights that signal processing steps to remove noise improve the results of seizure detection algorithms.

*Reducing computational costs.* Feature selection methods play a vital role in reducing computational complexity, improving computing times, and enhancing accuracy. Savadkoohi et al. used T-test and Sequential Forward Floating Selection (SFFS) to select significant features from EEG signals, achieving a classification accuracy exceeding 99.5% [38]. Tran et al. employed the discrete wavelet transform and a binary particle swarm optimizer to reduce data dimensionality by 75% while achieving an accuracy of 98.4% and reducing the computational time by 47% [39]. Atal et al. combined a modified Blackman bandpass filter-greedy particle swarm optimization (MBBF-GPSO) and CNN to achieve a seizure classification accuracy of 99.65% [40]. Through proper data analysis techniques, it is possible to extract relevant features from the EEG signals which can significantly reduce computational costs for a more optimized detection method.

Several approaches and commercialized products have emerged using signals from alternative sources such as Electrocardiography (ECG) and Photoplethysmography (PPG) [41–44], Electromyography (EMG) [45–49], or even Electrodermal Activity (EDA) [7, 50, 51] in a range of form factors. Combining multiple physiological signals such as EMG, ECG, EOG, motion, as well as audio/video recordings, boosts the accuracy of seizure detection [8]. Szabó et al. [52] utilized electromyography (EMG) to detect seizures, achieving high sensitivity and specificity. The works done in [53, 54] utilized electrocardiography (ECG) and heart rate variability (HRV) analysis to successfully detect seizure events. Our work focuses on EEG signals, which are commonly used in clinical settings.

However, there remains a significant gap in monitoring patients both before and after hospitalization. There are currently no devices available that are medically verified, have a convenient design, and have a socially acceptable form factor, that can provide patients with seamless and continuous monitoring before they get admitted to a hospital and after they are discharged. This limitation highlights the urgent need for innovative solutions that can ensure comprehensive seizure tracking and management in all settings.

## 2.2 Wearable-based Approaches

Wearable EEG devices are key to extending seizure studies beyond the hospital settings [55, 56]. Researchers have developed wearable EEG recording devices that can be paired with smartphones to continuously monitor a patient's EEG signals for later evaluation by doctors [57, 58]. Titgemeyer et al. [59] investigated the usability of wearable EEG devices by comparing them with vEEG signals. EEG recordings were simultaneously captured using the Emotiv EPOC [11] and a clinical vEEG system and performed a blind test with 10 independent raters, asking them to examine the recordings for abnormalities. They found that the wearable system had lower sensitivity and specificity compared to vEEG and concluded that wearable systems cannot yet replace classical EEG examinations. However, the lack of convenience with existing designs highlights the need for smaller and less cumbersome mobile EEG systems [60, 61]. Several companies are also exploring the feasibility of developing wearable systems as alternatives to clinical setups [12, 62, 63]. Wirst-worn devices like Empatica Embrace 2 [7] are unable to capture brain signals due to their placement away from the head. Conversely, head-worn devices such as [64, 65] are either too unwieldy or obtrusive for regular use while limitations such as battery life also restrict their continuous use. Finally, most of the current devices primarily focus on detecting one type of seizure, and including other seizure types lowers sensitivity and raises false detection rates [66]. The lack of convenience with existing designs highlights the need for smaller and less cumbersome mobile EEG systems.

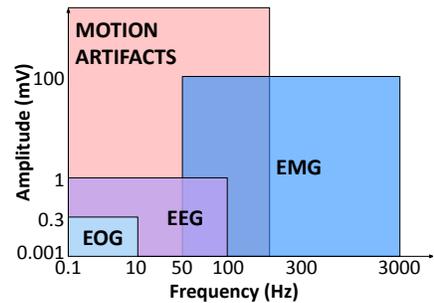

Fig. 3. The frequency and amplitude characteristics of ear signals and motion artifacts.

Ear-worn devices offer a promising alternative to traditional EEG systems for seizure detection, offering several advantages over scalp EEG and other wearable types [67–75]. Wireless, ear-worn devices are less cumbersome and more socially acceptable, increasing patient compliance for long-term monitoring. The absence of wires in wireless systems reduces noise degradation due to electrode wire movement [76]. They also enable easier and less intrusive data collection, making monitoring in non-clinical environments feasible. Unfortunately, the limited space inside can introduce difficulties in maintaining stable electrode contact and consistent signal quality. Moreover, EEG measurement requires differential measurements and the distance between electrodes needs to be at least 2 cm [77], otherwise the closely located electrodes could lead to electrical bridges and this is difficult to achieve in the limited space inside the canal. Developing comfortable and effective electrode designs for long-term wear without causing irritation or discomfort can also be difficult. In addition, Gu et al. [78] and You et al. [79] proposed an automated seizure detection system using electrodes placed behind the ear, confirming the feasibility of the idea. However, these studies did not thoroughly analyze the impact of motion artifacts, making their solution might not be ready for practical use. Even small artifacts such as EOG signal can cause poor performance [78]. Hence, more thorough studies on fully integrated systems in real-world clinical settings are necessary to validate the usability of the solution.

Fig. 4. International League Against Epilepsy (ILAE) 2017 Classification of Seizure Types Expanded Version. Red-marked types can be detected by *EarSD* due to the EEG, EOG, EMG sensors.

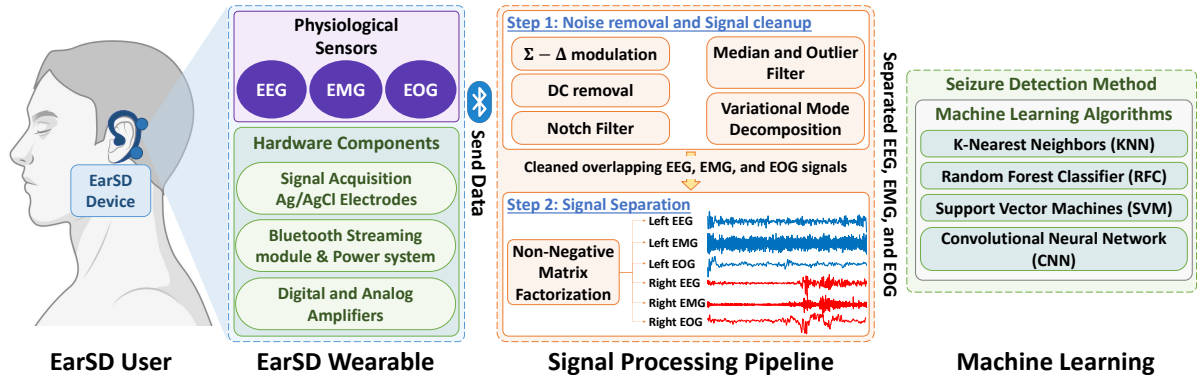

Fig. 5. *EarSD*'s system overview. The electrodes capture EEG, EMG, and EOG signals, wirelessly transmitting data to a host computer for noise filtering and signal separation. The decomposed signals are then analyzed by machine learning algorithms for automated seizure detection.

Moreover, epileptic seizure signals exhibit distinct characteristics in EEG recordings, including specific frequency bands, high amplitude spikes and sharp waves, and sudden onset and offset. They often show rhythmic discharges during the ictal period, with focal seizures displaying localized high-frequency activity and generalized seizures presenting spike-and-wave complexes. The signals can vary in spatial distribution, duration, and morphology, with interictal epileptiform discharges occurring between seizures. Accurate detection is challenging due to artifacts, the complexity of seizure patterns, and low amplitude signals, which can sometimes be lower than the typical amplitude ranges that wearable EEG devices are optimized to detect. As shown in Figure 3, the amplitude of EEG signal is much lower than the motion artifacts and other physiological types, such as EMG or EOG, making it hard to detect and recognize, let alone identify its seizure-associated signatures [80–83]

Last but not least, while vEEG systems can identify various seizure types (shown in Figure 4 [84, 85]), current wearable devices like Empatica are limited to detecting only tonic-clonic seizures due to their reliance on only wrist-based signals. While there might be a possibility to detect highlighted seizure types using ear-worn systems in Figure 4 because those types are what are recognized by vEEG, there have been limited studies that confirm the hypothesis.

## 3 *EARSD* SYSTEM

*EarSD* is a low-cost, unobtrusive, and comfortable ear-worn system designed with commercial-off-the-shelf (COTS) components to continuously monitor critical physiological signals associated with epileptic seizure onset. *EarSD* utilizes a non-invasive approach by capturing EEG, EMG, and EOG data from the upper and lower areas of the ears. The signals are wirelessly transmitted via Bluetooth to a host computer for further processing and analysis to detect if the user is experiencing a seizure. *EarSD* is equipped with EEG, EOG, and EMG sensors, enabling it to detect a broader range of seizure types, particularly those involving eye and muscle activity. This multi-modal approach offers significant advantages over commercially available devices, which primarily use EEG sensors. Unlike some studies that consider EOG and EMG signals as noise, our device leverages these additional data streams to enhance seizure detection accuracy. By combining data from EEG, EOG, and EMG, our device can better differentiate between actual seizure activity and non-seizure related movements, leading to more effective noise reduction and fewer false positives and negatives. This comprehensive approach offers a more thorough monitoring of individuals with epilepsy.

The core components of the *EarSD* include (1) an ear-worn sensing device, (2) a signal processing/decomposition pipeline, and (3) ML algorithms for seizure detection.

### 3.1 *EarSD* Hardware and Firmware

*Sensing Hardware.* The sensing hardware consists of two primary components: a brain-computer interface (BCI) and a pair of biosensor stickers (the wearable device). The first component, the BCI, featuring an ultra-low noise analog front-end (TI ADS1299) and an energy-efficient ARM Cortex-M4 microcontroller (TI MSP432P401), utilizes ultra-low noise amplifiers and a 24-bit ADC chip for signal digitization. The first-order analog filters remove high-frequency noise from the environment before passing it to the low-noise differential amplifiers of the ADC. The ADS1299 also contains an integrated second-order $\Sigma - \Delta$ modulator that samples the input at 1.024 MHz and shapes noise across the Nyquist bandwidth (i.e. 0-512 kHz). A third-order low-pass sinc filter then removes most of the noise at high frequency before decimation to 250 Hz for Bluetooth streaming to the host computer. The main processing unit, the MSP432 microcontroller, is responsible for controlling the analog front end, dynamically adjusting amplifier gain, and streaming data to the host device.

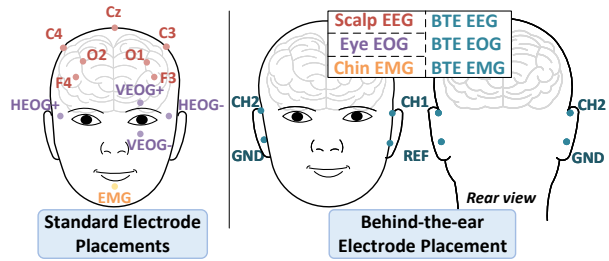

Fig. 6. *EarSD* has only 2 electrodes placed behind each ear whereas the standard placement has over 20 electrodes placed all over the head, eyes, and chin.

The second component is the biosensor electrodes, embedded on a pair of stickers that are fixed to the skin using disposable, double-sided adhesives behind the left and right ear. This design allows unobtrusive, continuous monitoring of the patient's bioelectrical signals without requiring an invasive or extensive setup. We use Ag/AgCl electrodes in our device as it is less prone to oxidation than other types of electrodes and thus ensure better accuracy and reliability of the captured signals. Each electrode is strategically placed, with two positioned symmetrically on the upper left and right ears while the other two serve as reference and bias electrodes located on the bottom of the left and right ears, respectively (Figure 6). The electrodes are placed close to the eyes, facial muscles, and regions of the brain, facilitating the recording of eye movements (EOG), muscle contractions (EMG), and mid-brain activity (EEG). Since EEG, EOG, and EMG are biopotential signals, they can be captured using the same electrodes. The electrode contact quality was regularly monitored so that we could detect and remove noisy signals created by loose electrodes. We also minimized electrical noise from the connecting wires by shielding them with two layers of aluminum and plastic. They are further shielded inside the stickers when connected to the electrodes, ensuring the subject's safety and preventing direct contact with the connections on their skin.

*Real-time Acquisition Software.* The real-time data acquisition firmware controls the operations and collects the physiological data measured from behind the ears through Bluetooth. It can be deployed on a laptop or a smartphone. The system produces low electrical risks as the electricity supplied to the sensing hardware is from a 3.7V rechargeable Lithium-Polymer battery. The receiver and batteries are enclosed in an electrically inert cover and casing. The sensing hardware and the battery power supply are enclosed in a small and lightweight plastic box to increase the subject's safety while using the system during the study.

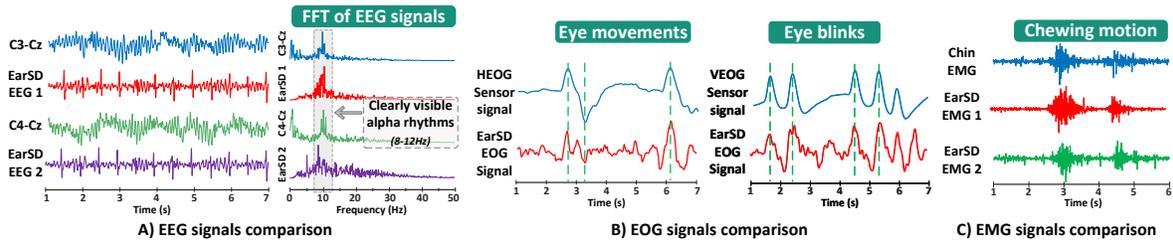

Fig. 7. *EarSD* EEG, EOG, and EMG signals compared to signals captured by dedicated sensors. The outputs show strong similarity between the dedicated sensors and the *EarSD* sensors for all three modalities validating the reliability of the signals recorded by *EarSD*.

### 3.2 In-lab Validation

We conducted experiments to verify *EarSD*'s ability to capture EEG, EMG, and EOG signals from behind-the-ear electrodes. Measurements were compared against ground-truth sensors positioned on the scalp, chin, and eye as per the standard International 10-20 system as illustrated in Figure 6. The ground truth EEG, EOG, and EMG signals were acquired using an FDA-approved Lifeline Trackit Mark III device. Data was acquired for one hour with the subject seated and the resultant signals obtained from both *EarSD* and the ground-truth sensors are visualized in Figure 7. In this setup, electrodes were placed side-by-side with the Standard Electrode Placements at the same positions. This was done to ensure that both sets of electrodes were capturing signals from the same location as closely as possible to validate that our device was recording accurate signals. Figure 7a represents the Alpha ($\alpha$) rhythms seen on both ear electrodes for EEG when the eyes are closed. $\alpha$-rhythms are prominent electrical oscillations in the frequency range of 8 to 12 Hz. Their presence indicates the sensor's ability to detect the subtle electrical activities of the brain and proves that *EarSD* can discern specific frequency bands of different brain states which are needed for capturing EEG patterns. Similarly, Figures 7b and c show similar EOG and EMG outputs between the dedicated sensors and *EarSD* for actions such as eye blinks, left-right eye movements, and chewing motion. Thus this confirms that *EarSD* can capture the important EEG, EMG, and EOG signals for seizure detection.

Currently, our system is used alongside the hospital's vEEG device as we shall describe in Section 5. This comparative approach helps us ensure that *EarSD* provides reliable data. Once *EarSD* is fully validated and optimized, we can deploy it as a stand-alone device in the real world that can complement and augment the hospital data so that doctors can have access to continuous and reliable recordings from all types of environments even when they leave the hospital.

## 4 SIGNAL PROCESSING

In this section, we highlight our approach to mitigate the motion artifacts and decompose the signals of interest.

### 4.1 Root Cause of Motion Artifacts

In order to ensure the reliability and practicality of *EarSD*, it is crucial to address the issue of artifacts contaminating physiological signals. Fundamentally, electrodes capture voltage differences across skin surfaces created by electric currents from brain tissue (as EEG), muscle (as EMG), and eye polarity (as EOG). These currents go through multiple layers, including the skull, skin layers (Stratum Corneum), and conductive gel to reach the electrodes [86]. The layers can be electrically modeled as resistors and capacitors as shown in Figure 8. While the skull has stable electrical properties, the skin, and conductive gel show electrical variations that introduce noise and artifacts caused by electrode inertia during activities like walking, shaking, or other repetitive activities [87]. Seizure events are particularly problematic due to their associated symptoms such as uncontrolled muscle spasms, which introduce significant motion artifacts. The relationship between inertial acceleration and voltage variation is non-linear due to the complex variation of electrical properties of skin layers.

In our proposed *EarSD* device, motion artifacts are even more severe for two key reasons. First, as a mobile device designed for daily use, *EarSD* is more susceptible to motion than ambulatory devices. Second, we prioritized user comfort in our design which means it was not possible to use glue to fix the electrodes to the skin. This compromises the stability of the skin-electrode interface. Hence, having robust noise removal techniques is necessary to minimize their impact on the seizure detection algorithm.

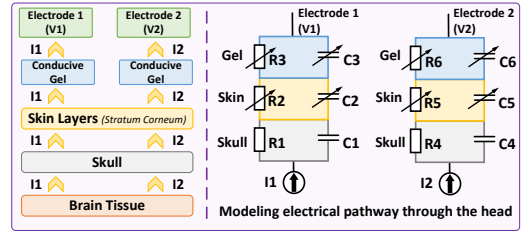

Fig. 8. Electric currents pass through the skull, skin, and the conductive gel to reach the electrodes. The skull is modeled as a fixed resistance path, while the skin and gel have variable resistance and capacitance.

As illustrated in Table 1, the overlapping frequency ranges and significantly higher amplitudes of motion artifacts compared to physiological signals such as EEG and EOG present substantial challenges in analyzing real EEG signals. EOG signals, which are related to eye movements, occupy a frequency range of approximately 0.1 to 10 Hz with amplitudes ranging from 0.001 mV to 0.3 mV. EEG signals, reflecting brain activity, span a broader frequency range from 0.1 to 100 Hz but typically maintain amplitude levels of less than 1 mV. EMG signals, generated by muscle activity, are the strongest of the signals, covering a wide frequency range from 50 to 3000 Hz and exhibits much higher amplitudes, reaching up to 100 mV. In contrast, motion artifacts, which is the unwanted portion of the signal analysis step, affect a wide frequency range (0.1 to 100 Hz) and can reach amplitudes higher than even 100 mV [80–83]. The overlap and the higher amplitude of motion artifacts complicate the task of accurately isolating and analyzing the true physiological signals. Effective noise reduction and artifact removal techniques are essential to mitigate these challenges and ensure reliable EEG signal analysis, particularly for applications requiring precise detection of brain activity such as seizure monitoring.

The ADS1299, used in *EarSD*, features an integrated second-order Sigma-Delta ($\Sigma-\Delta$) modulator, which significantly enhances signal quality through advanced noise reduction. By oversampling the input signal, the quantization noise was spread over a wider frequency range, reducing its impact within the low-frequency bands of interest. The noise shaping function of the $\Sigma-\Delta$ modulator further pushed the quantization noise to higher frequencies. After this process, digital filtering attenuates the high-frequency noise, ensuring that the low-frequency signals remain clear and accurate. The $\Sigma-\Delta$ modulation effectively removed noise from the low-frequency physiological signals, enhancing the signal-to-noise ratio (SNR) and enabling us to make high-fidelity recordings. All recorded data is passed through a standard notch filter to eliminate power line interference at 50/60 Hz. Linear trends are removed to prevent DC drift effects, and an outlier filter is applied to exclude transient spikes and ripples.

| Captured Signals | | Frequency | Amplitude |
|---|---|---|---|
| **EEG** | Delta ($\delta$) | <3 Hz | <1mV |
| | Theta ($\theta$) | 3-8 Hz | |
| | Alpha ($\alpha$) | 8-12 Hz | |
| | Beta ($\beta$) | 12-25 Hz | |
| | Gamma ($\gamma$) | >25 Hz | |
| **EOG** | | 0.3-10 Hz | 0.001-0.3 mV |
| **EMG** | | 10-100 Hz | <100 mV |

Table 1. Characteristics of the signals captured by *EarSD*

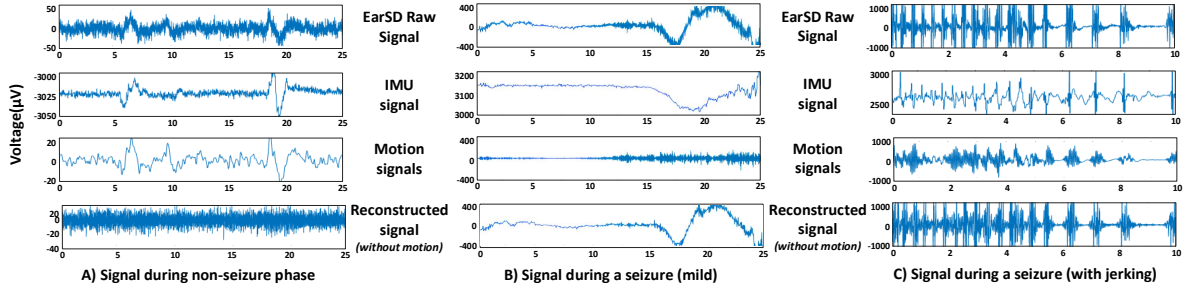

Fig. 9. VMD yields IMFs with distinct correlation patterns to the IMU data. Selectively reconstructing using IMFs uncorrelated with motion data filters motion artifacts removing noise while retaining critical seizure waveforms.

## 4.2 *EarSD* Software-based Motion Removal.

We found motion artifacts to span all frequencies of interest with high unpredictability, making their removal challenging through filtering or Independent Component Analysis (ICA) [88]. It is important to note that our data was collected in a hospital setting, which is much more ideal than the home setting. Even under such controlled conditions, we found noise to be a consistent feature. This problem will only be exacerbated when the device is deployed in the real world. While Active Electrodes (AE) have been proposed to mitigate motion artifacts [89], conventional designs do not consider behind-the-ear signals which are weak, overlapping, and constrained by limited space. Hence, we implement tailored measures to address the noises.

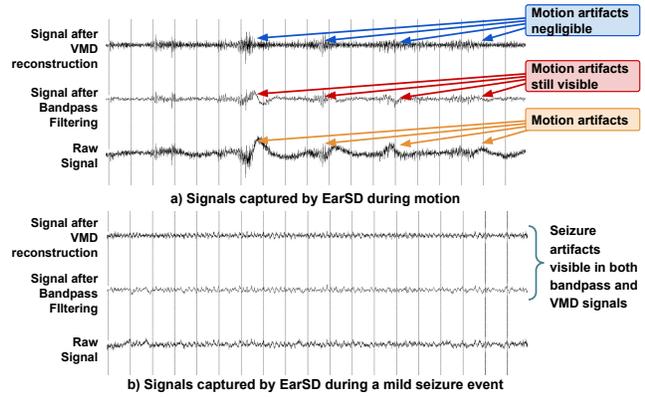

Fig. 10. Comparison of bandpass filtering with signal reconstruction using VMD

We employed Variational Mode Decomposition (VMD) to decompose the physiological signal into various components called Intrinsic Mode Functions (IMFs) [90]. IMFs are Amplitude Modulated-Frequency Modulated (AM-FM) signals $u_k(t) = A_k(t)cos(\phi_k(t))$ where $\phi_k(t)$ is the phase (non-decreasing function) and $A_k(t)$ is the non-negative envelope. The VMD process produces IMFs with distinct correlation patterns to IMU signals, facilitating selective reconstruction. Then, using data gathered concurrently from the Inertial Measurement Unit (IMU) of the device, the acceleration amplitude is calculated and compared with the motion artifacts. Subsequent analysis of IMU data revealed that the amplitude of acceleration correlates with motion artifacts.

The VMD process yields several IMFs with distinct correlation patterns to the IMU signals, allowing for selective reconstruction of the physiological signal. By computing correlations between physiological IMFs and IMU data, we can identify motion-related components for selective reconstruction, excluding any distortions due to noise. The reconstructed signal post-VMD showcasing the mitigation of motion artifacts is illustrated by Figure 9a. As can be seen from the figure, most of the motion-related events are removed from the signal. The same technique can be applied to remove most of the motion artifacts caused during a seizure (Figures 9b and c).

To evaluate VMD's effectiveness, we also compared the reconstructed signal with a [1 30] Hz bandpass filter. In Figure 10a, we can see that the 1 – 30 Hz filter cannot fully remove the motion artifact from the raw signal, while the VMD reconstructed signal has no motion artifacts Furthermore, during seizure events, VMD reconstructed signal still preserves the rhythmic slow wave artifacts and is comparable to the filter technique (Figure 10b). Therefore, we observe that VMD can retain seizure-characteristic information while excluding motion-induced distortions. Finally, we evaluated the impact of motion artifacts on our seizure detection algorithms by running them on our dataset, once with motion and then again with the motion artifacts removed. We discuss the results in detail in Section 6.2.3.

### 4.3 Decomposing the Denoised Signals into EEG, EMG, and EOG Constituents

Combining EEG, EMG, and EOG in the algorithm helps reduce false positives in seizure detection algorithms by providing a more complete picture of neurological activity. The ML algorithms can learn to distinguish seizure *signatures* spanning the modalities that are distinct from normal traits present in each signal. Further, training on multidimensional input can better discriminate artifacts from neurological phenomena, and separating the signals is an important step in the process. Table 1 shows the frequency and amplitude ranges of the key biosignals acquired by *EarSD*, specifically 3-25 Hz/1mV for EEG, 0.3-10 Hz/0.001-0.3mV for EOG, and 10-100 Hz/100mV for EMG. However, it is challenging to separate the low-amplitude EEG and EOG signals over-lapped with high-amplitude EMG signals. To overcome this, we investigated various signal decomposition techniques commonly utilized in seizure detection applications. Standard EEG data analysis often uses filtering methods [91, 92], which have limited efficacy when signals overlap. Alternative approaches such as Independent Component Analysis (ICA) [93, 94] and Principal Component Analysis (PCA) [95, 96], typically presuppose signal independence which is not always met by physiological signals. Through our analysis, we determined that Empirical Mode Decomposition (EMD) effectively separates the composite signal into distinct components with unique frequency resolutions. These components, when properly combined, facilitate the accurate reconstruction of the original signals. Additionally, we evaluated Non-Negative Matrix Factorization (NNMF), which leverages pre-trained frequency templates for signal differentiation. EMD and NNMF were chosen due to their advanced capability in signal separation tasks.

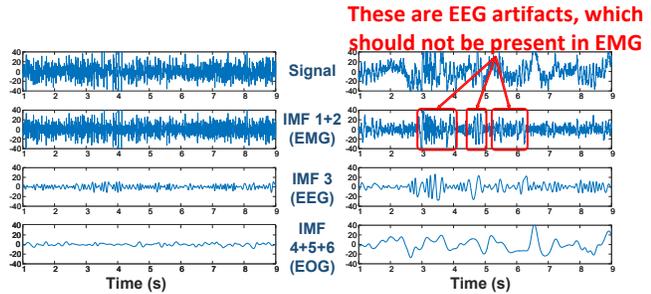

Fig. 11. EMD IMFs 1 and 2 typically capture EMG activity, IMF 3 aligns with EEG, and IMFs 4-6 show EOG, but IMF 2 for EMG is inconsistent, sometimes including EEG artifacts due to the overlapping nature.

*4.3.1 Empirical Mode Decomposition (EMD).* EMD is a robust technique for analyzing non-linear and non-stationary data by decomposing a signal into its Intrinsic Mode Functions (IMFs). This facilitates detailed time-frequency analysis while retaining the data in the time domain [97, 98]. IMFs exhibit three key properties: (1) Each IMF represents a single frequency at any given time, enabling multiresolution decomposition of the composite signal. (2) The average value of the oscillatory components within each IMF is zero. (3) The IMFs are mathematically orthogonal to one another.

By correlating IMF frequencies with known EEG, EOG, and EMG ranges across two separate data segments of the same patient, we observed that IMFs 1 and 2 typically capture EMG activity, while IMF 3 aligns with EEG, and IMFs 4, 5, and 6 with EOG. However, as Figure 11 shows, the assignment of IMF 2 to EMG is not consistent, as the range of an IMF is contingent upon the frequency content present in the mixed signal. Through our analysis, we determined that assigning IMF 1 for EMG, IMF 3 for EEG, and IMFs 4 to 6 for EOG yielded more accurate results.

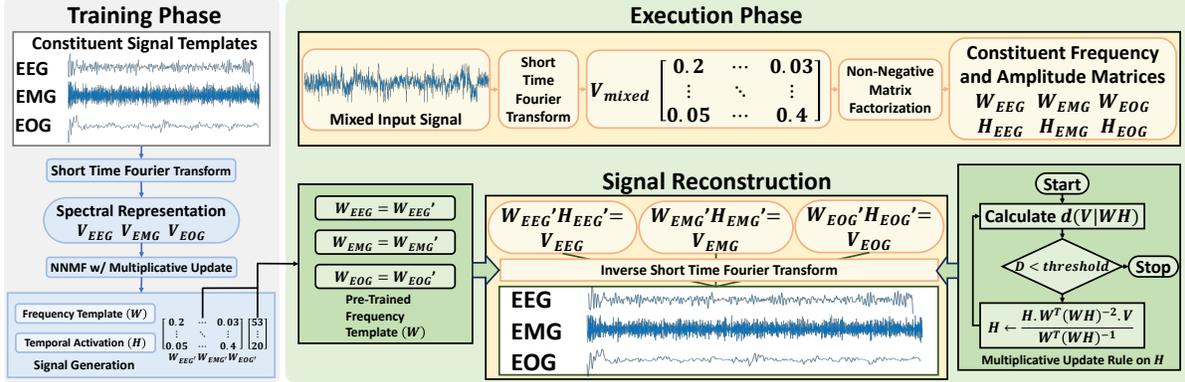

Fig. 12. Overview of supervised NNMF signal separation algorithm, Using templates of known EEG, EOG, and EMG signals in the training phase, NNMF can decompose a mixed physiological signal through an iterative process that minimizes divergence between the original and generated signals for accurate reconstruction in the execution phase.

*4.3.2 Non-Negative Matrix Factorization (NNMF).* NNMF factors a non-negative matrix into two lower-dimension matrices through multiplication [99]. The equation is given by $V = W * H$ where $V$ is the original non-negative matrix, $W$ is the frequency template matrix), and $H$ is the activation matrix. NNMF has various applications, including dimension reduction [100, 101], feature extraction [102, 103], and blind source separation [104, 105] making it suitable for our purpose. In signal processing, NNMF is particularly useful for disentangling one-dimensional signals by leveraging the non-negative properties of their spectral representations [106–108]. If $V$ is a spectral representation of a signal, its factorization $W$ would be considered as the frequency template and $H$ is the temporal activation of the signal. That is to say, $W$ will represent for the frequency inside the signal and will not change for EOG, EEG, and EMG and we can decompose and reconstruct the signal accordingly if we know the frequency template of EOG, EEG, and EMG. Supervised NNMF-based separation algorithm utilizes known physiological signals to train the model, then applies that model to signal separation and reconstruction [109]. Figure 12 shows the overview of our supervised NNMF approach.

**Training Phase.** During training, signal-specific templates are extracted from channels known to be dominated by each modality to derive the frequency basis matrix $W$. EEG segments are extracted from the C3-P3 and C4-P4 channels. EMG artifacts are typically pronounced in channels P7-TP9 and P8-TP10, so this is chosen as the EMG template, and channels FP9 and FT10, known for capturing EOG artifacts like blinks and saccades, are used to obtain EOG templates. We also used signals during well-studied seizure events for each of the three modalities to ensure that our signal templates included traces from both seizure and non-seizure periods.

Once the EEG, EMG, and EOG templates are acquired, NNMF with multiplicative updates is applied to construct the frequency template $W$. The updates progressively refine $W$ and the activation matrix $H$ through an iterative process, minimizing the divergence between matrices [110]. The error in the factorization process necessitates a metric for the distance between two matrices. One well-established measure is the $\beta$ divergence [111] with three common variations defined as:

$$d_\beta(X|Y) = \begin{cases} \frac{1}{\beta(\beta-1)}(X^\beta + (\beta-1)Y^\beta - \beta XY^{\beta-1}) & \beta \in \mathbb{R}\{0,1\} \\ X\log\frac{X}{Y} + (Y-X) & \beta = 1 \\ \frac{X}{Y} - \log\frac{X}{Y} - 1 & \beta = 0 \end{cases}$$

The $\beta$ divergence includes three commonly utilized variations [112]:

- Euclidean divergence ($\beta = 2$): $d_{EUC}(X,Y) = \sqrt{\Sigma_{i,j}(X_{i,j} - Y_{i,j})^2}$
- Kullback-Leibler divergence ($\beta = 1$): $d_{KL}(X,Y) = \Sigma_{i,j}(X_{i,j}\log\frac{X_{i,j}}{Y_{i,j}} - X_{i,j} + Y_{i,j})$
- Itakura-Saito divergence ($\beta = 0$): $d_{IS}(X,Y) = \frac{X}{Y} - \log(\frac{X}{Y}) - 1$

One important characteristic that affects our approach is their scale invariance properties:

- $d_{EUC}(\lambda X|\lambda Y) = \lambda^2 d_{EUC}(X|Y)$
- $d_{KL}(\lambda X|\lambda Y) = \lambda d_{KL}(X|Y)$
- $d_{IS}(\lambda X|\lambda Y) = d_{IS}(X|Y)$

The Itakura-Saito divergence ($d_{IS}$), due to its scale-invariance, is particularly suitable for representing data with significant dynamic ranges, such as physiological signal spectra. NNMF with a multiplicative update rule for $d_{IS}$ divergence is then applied in the training phase.

$$H \leftarrow H \cdot \frac{W^T(WH)^{-2} \cdot V}{W^T(WH)^{-1}}; W \leftarrow W \cdot \frac{((WH)^{-2} \cdot V)H^T}{(WH)^{-1}H^T}$$

By leveraging known EEG, EMG, and EOG patterns containing both normal and epileptic traits, supervised NNMF can effectively decompose the composite ear signal for selective reconstruction of each modality.

**Execution Phase.** In the execution phase, mixed signals obtained from *EarSD* are separated using the frequency template $W$ derived in the training phase. Since the frequency template remains unchanged for each signal, we apply an STFT to each data segment to acquire its spectral form. The multiplicative update rule is subsequently employed on matrix $H$ to extract temporal activation of individual signal components within the mixed data. This process strives to minimize the distance between the reconstructed and the original signals, hence reducing the error inherent in the factorization. The final reconstructed signal is obtained by inverse STFT using the component matrices $W$ and $H$ once divergence is sufficiently minimized.

## 5 SEIZURE DETECTION USING *EARSD*: A REAL-WORLD STUDY

### 5.1 Study Protocol

To evaluate the effectiveness of the proposed *EarSD* device in seizure detection, we conducted a clinical study at the Epileptic Monitoring Unit (EMU) of a hospital. Our study aimed to demonstrate that *EarSD*'s performance is comparable to the "gold standard" of video EEG monitoring in hospitals. We required patients to wear both the *EarSD* device and standard 21-channel scalp-EEG with video recording simultaneously to ensure that both devices were collecting the same data from the same patient for the same times (Figure 13). This allowed us to compare the results and verify if the same events were detected by both.

*Patient Recruitments.* To evaluate the effectiveness of the proposed *EarSD* device in seizure detection, we conducted a clinical study at the Epilepsy Monitoring Unit (EMU) of UTSW Medical Campus hospital. Through our collaboration with the hospital, we gained access to patients admitted to the EMU for long-term vEEG monitoring.

To be eligible for our study, individuals had to be at least 18 years old at the time of enrollment and willing to wear the *EarSD* device. Patients wearing any other ear device, such as hearing aids, or intracranial electrodes, were excluded from the experiment. Anyone unable or unwilling to provide informed consent was also excluded. Following these rules, we were able to recruit 32 patients aged between 19 and 74, with 16 biological males and 16 biological females represented in the sample over 09 months. Description of the participants are listed in Table 2. Note that SD_015 was omitted from data analysis as this was the same patient as SD_020. This patient was enrolled twice and only had seizures during his second admission. As such, only the data from EarSD_020 were used for analysis and reporting.

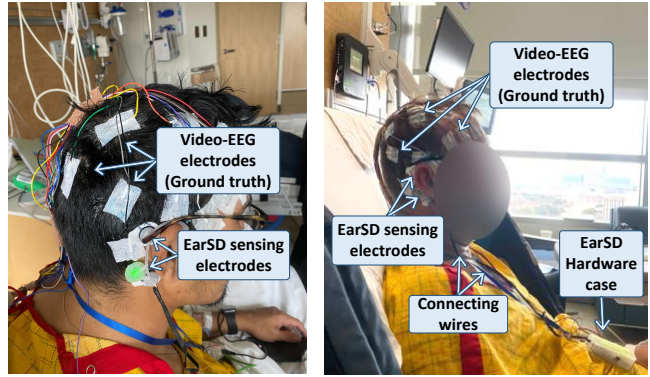

Fig. 13. A patient in the hospital wore the standard vEEG electrode and the *EarSD* wearable behind the ears at the same time for data collection.

| Patient ID | Age/Sex | Seizure Freq. | Duration (Hours) | Patient ID | Age/Sex | Seizure Freq. | Duration (Hours) |
|---|---|---|---|---|---|---|---|
| EarSD_001 | 74 M | ≤1/year | 72 | EarSD_018 | Unknown | ≥1/year | 24 |
| EarSD_002 | 51 M | ≥1/year | 0 | EarSD_019 | 36 M | ≥1/year | 48 |
| EarSD_003 | 53 M | ≥1/year | 48 | EarSD_020 | 26 M | ≥1/month | 96 |
| EarSD_004 | 55 F | ≤1/year | 48 | EarSD_021 | 53 F | ≤1/year | 24 |
| EarSD_005 | 38 M | ≥1/year | 24 | EarSD_022 | 65 F | ≥1/year | 24 |
| EarSD_006 | 44 F | ≥1/week | 12 | EarSD_023 | 73 F | ≥1/year | 24 |
| EarSD_007 | 40 M | ≤1/year | 48 | EarSD_024 | 24 F | ≥1/month | 24 |
| EarSD_008 | 49 F | ≤1/year | 12 | EarSD_025 | 38 F | ≥1/month | 96 |
| EarSD_009 | 40 M | ≥1/month | 24 | EarSD_026 | 53 F | ≥1/week | 24 |
| EarSD_010 | 49 F | ≥1/year | 48 | EarSD_027 | 35 F | ≥1/year | 48 |
| EarSD_011 | 46 M | ≤1/year | 72 | EarSD_028 | 63 M | ≥1/year | 24 |
| EarSD_012 | 19 M | ≥1/year | 48 | EarSD_029 | 21 F | Daily | 48 |
| EarSD_013 | 27 F | ≥1/year | 48 | EarSD_030 | 33 F | ≥1/year | 48 |
| EarSD_014 | 66 M | ≥1/week | 72 | EarSD_031 | 43 M | ≥1/year | 48 |
| EarSD_016 | 35 M | ≥1/year | 24 | EarSD_032 | 24 F | ≥1/year | 24 |
| EarSD_017 | 40 M | ≥1/year | 48 | EarSD_033 | 64 M | ≥1/week | 24 |

Table 2. Patient demographics and *EarSD* usage details. F = female; M = male.

*Data Collection Procedure.* At the start of the study, before the device placement, the area behind each ear was examined for any pre-existing skin condition that might have hampered the skin-electrode contact. After obtaining formal written consent, the subject wore *EarSD* and the standard 21-channel scalp-EEG with video recording (Natus NeuroWorks EEG Software [113]) simultaneously. The *EarSD* board, was placed around the patient's neck using a detachable lanyard, and the sticker electrodes were attached behind the patient's ears using collodion glue to ensure firm contact between the skin and the electrode. We ensured that the impedance of the EEG electrodes was regularly monitored and maintained within acceptable ranges (typically below 5 $k\Omega$ [114]) to ensure high-quality signal acquisition. We measured the impedance at each electrode regularly during our routine checks and adjustments during the data collection. Impedance was measured using the '*Lead Off Detection*' feature of the ADS1299. This technique involves injecting a tiny (6nA) current into the electrode, resulting in a voltage that can be measured. The high input impedance of the differential amplifier ensures that no current flows into the other electrode line, allowing for accurate impedance measurement. Therefore, we can easily measure just the voltage drop across the first three elements — the 5 $k\Omega$ in-series resistor, the electrode-to-skin impedance, and the impedance of a portion of the human body. Because the series resistor is known and the body's impedance is too small to matter, we have only one unknown remaining — the impedance of the electrode-to-skin interface. We modeled the electrode-to-skin interface as a simple resistor to estimate the result accurately. Using these values, we could then calculate the impedance (Z) using: $Z = \dfrac{V\sqrt{2}}{I} - 5k\Omega$ where $V$ is the measured voltage and $I$ is the known current. We measured the voltage as an RMS value, whereas the 6nA current is an amplitude, not an RMS value. So, our calculation included a factor of $\sqrt{2}$ to convert RMS into amplitude and we subtracted the 5kOhm to remove the built-in resistor of the board. This, combined with our regular impedance checks, enabled us to obtain and maintain high-fidelity EEG recordings.

Once the electrodes were placed comfortably over each ear, the sensing circuit was paired via Bluetooth to a tablet for real-time data storage and viewing. The participants were encouraged to wear our device for as long as they felt comfortable during their EMU stay, including during sleep. As per standard clinical protocol, they were then monitored in their rooms for the duration of their stay in the hospital. Among all participants, *EarSD* was worn for a total of 1250 hours with each patient wearing the device for an average of 41 hours. Given our device is an early prototype, such extended use gives us confidence that our device is robust for long periods. Additionally, we tested the device in various conditions as the patients always wore it, even sleeping with it, so we ensured that *EarSD* can consistently capture high-quality signals across different situations. Furthermore, consistently recording physiological signals from 32 patients, each using the device for 1 to 4 days on average, we are confident that our signal acquisition process is repeatable as we were successful in collecting their data over multiple sessions under similar conditions. Note that most of the patients enrolled in our study did not experience a seizure during their time in the hospital. Of the 32 participants, only 7 had a seizure attack which further highlights the need for continuous seizure monitoring devices that patients can wear outside of the hospital, as seizures happen at any time.

At the end of the evaluation, all information that was gathered was deidentified, except for the patient's study identification number. The data was then stored on an encrypted and password-protected laptop for processing before being uploaded to the REDCap (Research Electronic Data Capture) Database. Local data on the laptop was destroyed after uploading to REDCap. The data collected from subjects is unidentifiable and was only shared between the researchers participating in this study.

## 5.2 Data Preparation and Clinical Verification

After the data was processed using the signal processing method described in Section 4, we plotted the EEG signals from the EarSD device alongside hospital video-EEG recordings as shown in Figure 14, allowing our collaborating doctors to review the signals and annotate seizure onset and offset times on both the hospital EEG and EarSD EEG signals. Given that all three signals (EEG, EMG, and EOG) were recorded simultaneously, we could use the same timestamps to identify the EMG and EOG signals during seizure periods. The annotated onset and offset times on *EarSD* data provided us with a labeled dataset to train the supervised machine learning algorithms. Given that seizures occur only for short periods, using all the recorded data, which includes long non-seizure periods, would create a highly unbalanced dataset.

This imbalance would lead to ineffective ML models that might not produce meaningful outputs. To address this, we created a balanced dataset by including seizure signals, non-seizure signals from periods before and after the seizures, as well as non-seizure signals from periods of daily activities such as talking, eating, and walking. This approach ensured that the dataset had a more balanced dataset between seizure and non-seizure signals. This balance is crucial for training effective machine learning models, as it prevents the model from being biased towards non-seizure periods, thereby improving the reliability and accuracy of the seizure detection algorithm.

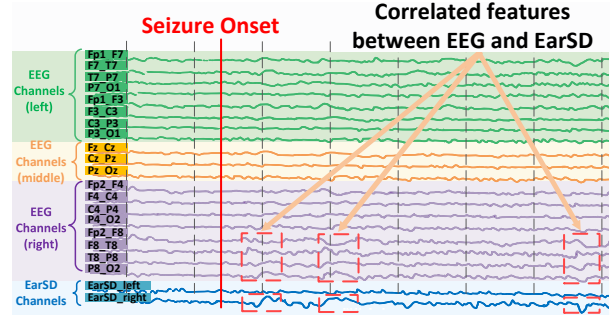

Fig. 14. Seizure event onset marked by the doctors shows a strong correlation between the vEEG and the *EarSD* channels. The spikes can be seen on the right channels of both devices, allowing *EarSD* to classify seizures as well.

Per American Clinical Neurophysiology Society guidelines, an abnormal event is considered a seizure if it lasts for at least 10 seconds [115]. Thus, we segmented the events into 10-second chunks for analysis. A chunk entirely within the labeled seizure onset and offset time was labeled as a seizure, while chunks partially overlapping or out of seizure onset and offset times were labeled as non-seizure. A subset of the dataset containing the onset and offset times is provided in Table 3.

### 5.3 *EarSD*'s Seizure Detection Algorithms

*5.3.1 Feature Extraction.* We used the dataset prepared as described above for our machine learning task. Before passing our signals to the classifiers, we performed feature extraction to aid the algorithms in their detection task. We used the signals in our dataset (2 EEG, 2 EMG, and 2 EOG for the left and right ears, respectively) and extracted features in the time and frequency domains. For the time domain, we computed statistical features, namely the mean, standard deviation, average deviation, skewness, kurtosis, lowest Value, highest Value, and the root mean square amplitude. We calculated the Mel-Frequency Cepstral Coefficients (MFCCs), widely used in time-series signal processing for the frequency domain. We extracted 10 MFCCs per signal segment, resulting in 58 features per channel. Combining frequency and time domain features, we obtained 348 features, normalized between 0 and 1. The concatenated signals, which included all three modalities, were then used to train our machine learning models for seizure detection.

*5.3.2 Traditional Machine Learning.* We chose to examine three classical machine learning algorithms - Support Vector Machine (SVM), K-Nearest Neighbor (KNN), and Random Forest Classifier (RFC). We determined that since our eventual goal is to perform machine learning directly on the device, we wanted our model that would be accurate but also have low size and computation requirements. Additionally, since seizure signals are rare in the overall recordings, our signals of interest were not very large. We had to work with a subset to balance out the seizure to non-seizure signals and get an unbiased machine learning model. We found traditional machine learning models to fulfill our requirements, providing good accuracy with low computational overheads.

SVMs perform classification by finding the optimal maximum margin hyperplane that separates the classes. We used a Gaussian radial basis function (*rbf*) kernel to enable the SVM model to capture non-linear relationships in the feature space. The *rbf* kernel maps the input data into a higher dimensional space where classes can be separated by a linear decision boundary. The kernel coefficient, $\gamma$, was optimized at 0.5 and the penalty term, *C*, at 20 through grid search cross-validation to balance model complexity, accuracy, and overfitting. Overall, the SVM identifies the maximum margin decision boundary that best separates seizure and non-seizure EEG recordings. KNN is a non-parametric algorithm that classifies data points based on the class of their nearest neighbors in the feature space. We computed the Euclidean distance between points and set the number of neighbors, $K$, to 5 based on empirical tuning. The seizure/non-seizure label of each data point was determined by a majority vote among its 5 nearest neighbors enabling local, neighborhood-based classification. RFC is an ensemble method that aggregates the predictions of multiple decision trees. Each tree is trained on a random subset of features, which enables robust predictions even with correlated features. The forest consisted of 10 trees, each with a maximum depth of 100 based on optimization experiments. The consensus seizure/non-seizure label predicted by the forest was taken as the final output. Randomness introduced through bagging and feature subspacing makes RFC resilient to overfitting.

Due to the size of the dataset, we opted to use a Leave-one-out cross validation strategy as it allows us to get a more robust model that is less prone to overfitting and is more generalizable. The model was trained on all samples from the full dataset except one held-out patient in each fold. All data from this patient was kept separate from the test set and only used to test the model performance. We obtained a rigorous estimate of the model's ability to generalize to new patients by iterating through folds where each patient serves as the test set once and used the average accuracy from these runs to report the final accuracy. The same method was followed to test all three machine learning models (SVM, KNN, and RFC).

To avoid data leakage, we used the *scikit-learn* package. For each fold in the leave-one-out cross validation step, normalization parameters were computer exclusively from the training data of that fold, ensuring that the test data remained unseen during the calculation of these parameters. Specifically, we used '*StandardScaler*' function to calculate the mean and standard deviation from the training samples, applying the '*fit_transform*' method to the training data and the '*transform*' method to the held-out test data. This ensured that the test data was normalized using parameters derived solely from the training data, maintaining a strict separation between training and test sets.

*5.3.3 Neural Network Model.* We have also developed a Convolutional Neural Network (CNN) model for the seizure detection task. A key advantage of the CNN approach over traditional machine learning is that they can automatically learn features from the data for which they can pick up more complex features that we may have missed when manually selecting the features.

We trained a 1D CNN network consisting of three 1D convolutional layers with 32, 64, and 128, filters respectively. The kernel size was kept to 3 with padding and stride both set to 1 for all layers. We applied the ReLU activation function and used MaxPooling with a kernel size of 2 after each convolution. The convolutional layers were followed by three fully connected layers with 128, 128, and 64 units respectively. We also used a dropout layer with a probability of 0.5 after the first and second fully connected layers. The final fully connected layer outputs the binary classification outcome. We trained the model using the Adam optimizer and a weighted loss function (LF):

$$LF = \alpha \cdot (1 - p_t)^\gamma \cdot -log(p_t)$$

where $\alpha$ is the weighting factor for each class and is set to be inversely proportional to the class frequencies, $\gamma$ is the focusing parameter that controls the weighting of hard-to-classify samples, and $p_t$ is the model's estimated probability for the target class. This loss function approach addresses the issue of data imbalance by putting more focus on hard-to-classify samples, thereby enhancing model performance and reducing class bias. We trained the model for 350 epochs using 80% of the data for training and 20% for testing. Out of the training set, we use 20% as the validation set.

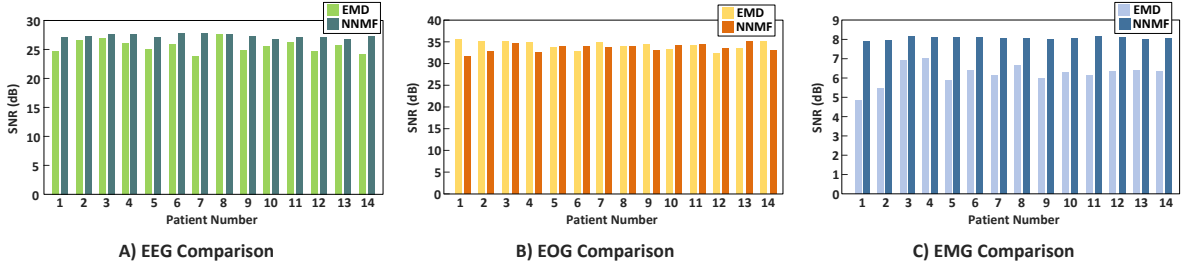

Fig. 15. Signal-to-Noise Ratio (SNR) between EMD and NNMF. The NNMF algorithm demonstrated superior EEG and EMG signal quality compared to EMD, with average SNR improvements of 1.64dB and 1.84dB respectively. For EOG signals, both NNMF and EMD yielded approximately equivalent SNRs.

## 6 PERFORMANCE EVALUATION

### 6.1 Sensitivity Analysis

*6.1.1 Noise Removal.* To demonstrate the effectiveness of the developed EMD-based and NNMF-based signal separation algorithms, we performed a thorough validation using data captured from the *EarSD* device in a clinical setting. For each patient, we extracted 2-hours of data - one hour during the daytime (between 1 p.m. to 3 p.m.) and one hour during the night (between 10 p.m. to 4 a.m.). The two-hour (7,200

| Patient ID | Start Time | End Time | Type |
|---|---|---|---|
| SD_004 | 21:42:56 | 21:43:21 | Focal Right |
| SD_005 | 01:01:01 | 01:01:58 | Generalized |
| SD_016 | 15:01:49 | 15:02:55 | Focal Left |
| SD_017 | 22:55:36 | 22:56:53 | Focal Left |
| SD_020 | 14:37:08 | 14:38:04 | Focal Left |
| ... | ... | ... | ... |

Table 3. Seizure log labeled by doctors

seconds) recordings are segmented into 720 10-second epochs to enhance computational efficiency.

Signal-to-noise ratio (SNR) is a well-known metric in digital signal processing, quantifying the target signal strength relative to noise. We estimated SNR based on the known frequency ranges of EEG, EOG, and EMG components. SNR for a signal within the frequency range $[a\ b]$ Hz was calculated as $SNR_{[a\ b]} = \frac{P_{[a\ b]}}{P_{other}}$ where $P_{[a\ b]}$ is the mean power of the signal in the frequency band of $[a\ b]$ Hz and $P_{other}$ is the mean power outside this band. After computing SNRs for each epoch, we determined the average SNR per patient shown in Figure 15. We can see that compared to the EMD-based algorithm, the NNMF-based algorithm demonstrated better performance in EEG and EMG signal quality, with comparable outcomes for EOG signals. Specifically, as can be in in Figure 15a, the average SNR values for EEG signals using the EMD-based algorithm ranged from 24.20dB to 27.66dB, with a mean of 25.69dB, while the NNMF-based algorithm ranged between 26.84dB and 27.77dB with a mean of 27.33dB. For the EOG signals, both algorithms yielded nearly equivalent average SNRs, 33.94dB for EMD and 33.77dB for NNMF as shown in Figure 15b. Figure 15c shows the results for the EMG with an EMD average SNR of 6.23dB, while NNMF averaged 8.07dB. So, we can conclude, that the NNMF approach provided an SNR enhancement of 1.64dB for EEG and 1.84dB for EMG signals over the EMD approach.

*6.1.2 Energy Consumption.* We measured the power consumption of the *EarSD* device using a Monsoon Power Monitor with a sampling rate of 5 kHz. Each measurement lasted 180s, resulting in 900,000 data points to get stable results. Under conditions of $250°C$ and a nominal battery voltage of 3.7V, the average power usage of our device was (1) Active state (sensing physiological signals, recording, and streaming via Bluetooth) consumed 241.5mW, (2) Idle state (MCU active with other components and streaming turned off) consumed 51.60mW.

With a 500mAh Li-Po battery, the *EarSD* device can operate for approximately 7.7 hours in the active state and 35.9 hours in the idle state. We also conducted component-level measurements during the active state by individually turning off each component and repeating the measurements. The sensing components (amplifiers and external ADCs) and the Bluetooth communication module were the primary power consumers, with an average power consumption of 93.5mW and 85.2mW, respectively, while the processing unit (MSP432) consumed only 62.8mW.

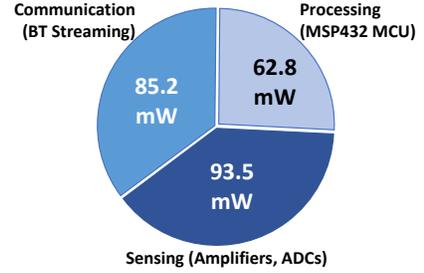

Fig. 16. Power Consumption of *EarSD*

These power consumption figures demonstrate the *EarSD* device's capability to monitor a user for extended periods. The current high energy consumption is due to engineering limitations, as direct power from the earbuds is not feasible and so requires a separate power source. However, optimization and implementation on a System-on-Chip (SoC) can enable power supply from the earbuds. By reducing the number of sensing components, optimizing Bluetooth transmission, and leveraging the MCU's deep power-saving modes, power consumption can be further reduced.

### 6.2 System Performance

The performance of the detection algorithms is evaluated over the dataset collected from our study at the hospital which contained all events from the patients who experienced seizures during our study. We test the performance of the machine learning algorithms (SVM, KNN, and RFC) through a leave-one-out strategy where we trained on all samples from the full dataset except one held-out patient which was used for testing. We also rotated the test sample so that all patients were tested in successive iterations. This approach showed the performance of the algorithms over specific events from each of our patients and can simulate the algorithm's performance in a real-world setting with unknown events.

#### 6.2.1 Identifying the Sliding Window Size.

We augmented our dataset using sliding windows. To evaluate the impact of sliding windows on the algorithm performance, we extracted the data with different overlapping windows. We varied it from 1 to 9 seconds to find the best configuration. With a 1-second sliding window, we slid the window 1 second forward which kept a 9-second overlap between the successive windows. Similarly, with a 2-second sliding window, the window was moved 2 seconds forward, which means there was an 8-second overlap between the two windows and so on. Using sliding windows also enabled us to capture spatial information and feed co-dependent data from earlier windows to the next window, helping the machine learning models improve their accuracy. It is evident from Figure 17a that all three algorithms show an accuracy of over 70% indicating their reliability in seizure detection using only data recorded from our *EarSD* device. We can see that when the stride size is set to 1, the results get vastly better, exceeding 90% accuracy for all three algorithms, with SVM achieving the best accuracy of 94.5%. In this configuration, there is maximum overlap between consecutive windows resulting in better training of the algorithms.

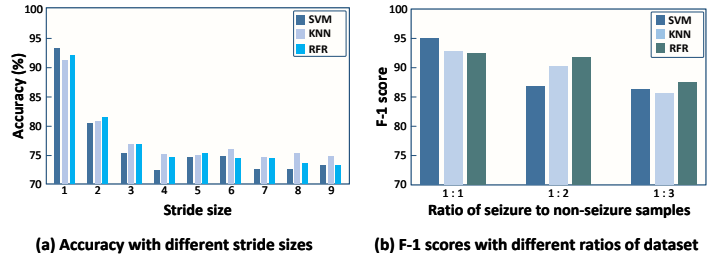

Fig. 17. All three algorithms show the best performance when there is maximum overlap between successive sliding windows and the dataset contains a 1:1 ratio between seizure and non-seizure samples.

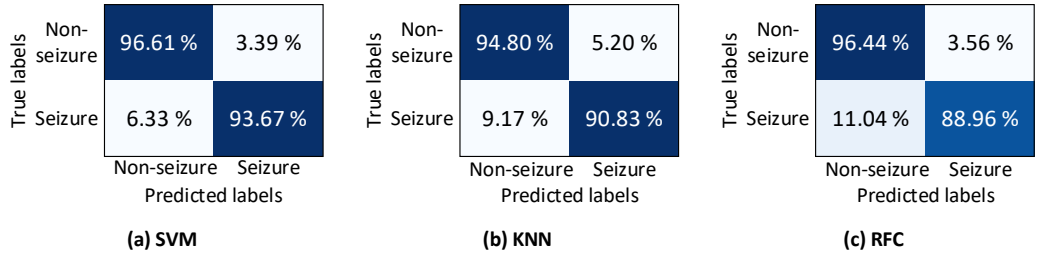

Fig. 18. Results of the seizure detection task. All three algorithms show accuracies of over 90% at seizure detection. SVM performs the best with an average accuracy of 95.3% at distinguishing between seizure and non-seizure events.

*6.2.2 Identifying the ratio of Seizure to Non-seizure samples in the dataset.* To investigate the impact of dataset bias, we calculated the F-1 scores by examining various ratios of seizure to non-seizure samples in the dataset. An imbalanced dataset leads to a decline in accuracy as the machine learning model becomes biased towards the majority class as we see in various other works [116, 117]. This disproportionality results in the algorithm producing false negatives making them unreliable. Our experimental results are depicted in Figure 17b which illustrates the F-1 scores achieved by the three algorithms when the ratio of seizure to non-seizure samples in the dataset are 1:1, 1:2, and 1:3. Even though all three algorithms attain high F-1 scores (exceeding 85%) indicating their capability in detecting seizures, the best results are obtained when there is an equal number of seizure samples to non-seizure samples with SVM showing the best F1-score of 95% in this case as well.

*6.2.3 Impact of Motion on Seizure Detection.* We investigated the effect of motion removal on the performance of the machine learning algorithms. We tested all three algorithms once on the dataset without removing motion and then again after motion is removed. For the SVM algorithm, accuracy increases from 84.3% without motion removal to 94.85% with motion artifact removal, showing a 10% improvement. The KNN classifier also sees an increase from 83% to 92%, demonstrating a similar improvement as SVM. Most significantly, RFC exhibits the most impact, with accuracy rising from 55% without motion removal to 93% with motion artifact removal, a substantial 40% improvement. Overall, the results illustrate that removing motion artifacts significantly boosts the accuracy of all classifiers, with SVM achieving the highest accuracy after motion artifact removal, and RFC showing the greatest sensitivity to such artifacts and the most dramatic improvement upon their removal.

Based on the conclusions drawn from these experiments, we perform our seizure detection task using a 1-second sliding window and keeping a 1:1 ratio between seizure and non-seizure samples in the dataset that has the motion artifacts removed.

*6.2.4 Seizure Onset Detection Performance.*
**Traditional Machine Learning:** The confusion matrices of Figure 18 show the performance of the ML algorithms by comparing the predicted labels with the actual labels. From this, we can assess the model's ability to correctly classify both positive (seizure) and negative (non-seizure) instances and evaluate its overall performance. From Figure 18a, we can see that the SVM algorithm is able to correctly detect seizures 93.67% of the time and non-seizures 96.61% of the time indicating its ability to correctly distinguish between seizure and non-seizure events. Similarly, Figure 18b, shows that the KNN algorithm achieves a seizure detection rate of 90.83% and a non-seizure detection rate of 94.80%. RFC also shows good performance in Figure 18c. We can see that it correctly detects seizures 88.96% of the time and non-seizures 96.44% of the time. We can thus conclude that SVM outperforms the other two algorithms showing the best overall results.

**Neural Network Method:** Neural networks offer some critical advantages over the traditional machine learning methods which makes their development an important step in our overall task of seizure detection. CNNs can automatically learn relevant features from raw data through multiple layers of convolutional filters allowing them to capture intricate patterns and representations in the data that might be missed with manual feature extraction. Traditional machine learning requires manual feature extraction, which involves domain expertise and can be time-consuming. The quality of the model depends heavily on the chosen features and this makes them unmanageable as the dataset increases. They are also dependent on the quality of features chosen which can affect the outcome of the models and deep learning methods are more robust in this respect. Therefore, we trained the CNN model on the same EEG dataset used for traditional machine learning models and present the results in Figure 20b.

As can be seen in the figure, the CNN model demonstrates superior performance compared to the SVM model (the best performing model of the traditional methods). The CNN model correctly identified 98.95% of non-seizure instances, outperforming the SVM's 96.61%, and significantly reduced the false positive rate from 3.39% to 1.05%. Additionally, the CNN model achieved a higher accuracy in identifying seizure instances, with a correct identification rate of 95.13% compared to the SVM's 93.67%. It also reduced the false negative rate from 6.33% to 4.87%, indicating improved sensitivity in detecting seizures. These enhancements underscore the benefits of using advanced machine learning techniques like CNNs, which can automatically learn and extract relevant features from the data, leading to better overall performance in complex tasks such as seizure detection.

The key performance metrics of all the machine learning algorithms that we investigated are presented in Figure 19. It shows the Precision, Recall, and F1 Score for both non-seizure and seizure events. Precision shows how many of the detected seizures were actual seizures, Recall shows what proportion of actual seizures was detected by *EarSD* and F1-score provides a balanced measure of the system's overall performance in detecting seizures correctly. For non-seizure events, CNN achieved the highest precision at 99%, followed closely by SVM and RFC at 97%. CNN also led in recall with 96%, outperforming SVM, KNN, and RFC. The highest F1 Score for non-seizure events was 98%, achieved by CNN. For seizure events, CNN again showed the highest precision at 98% and recall at 99%, indicating its superior performance in capturing true positive seizure events.

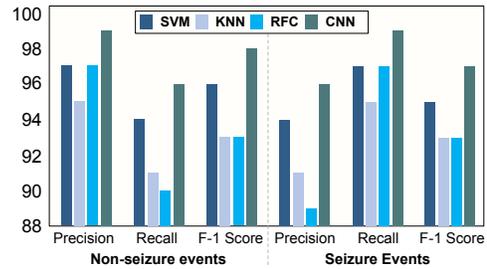

Fig. 19. Performance Metrics of all the Machine Learning Algorithms we investigated

In summary, the results show that with just two electrodes placed behind the ear, we are able to capture signals that can be used to reliably identify seizure events proving the effectiveness of our device. These promising results were achieved using our *EarSD* device, which captures EEG, EMG, and EOG data from two electrodes placed behind the ear, demonstrating the feasibility and effectiveness of our wearable device for continuous, non-invasive seizure monitoring. It should be noted that although *EarSD* has only been tested on a small number of patients due to limitations of funding, such high accuracy of detection is encouraging and approaches the standard of accuracy needed for medical devices to receive approval from regulatory bodies [118, 119].

*6.2.5 Comparative Study.* We further evaluated the performance of traditional machine learning and CNN methods for seizure detection using data from our proposed *EarSD* device and the hospital's 'gold-standard' vEEG system. As expected, when using only data from the vEEG device, our CNN network yielded the best results, achieving 99.44% accuracy for non-seizure detection and 96.61% for seizure detection, with minimal false positives (0.56%) and false negatives (3.39%) as shown in Figure 20c. The *EarSD* device, although slightly less accurate, showed impressive results with the CNN model, indicating that the wearable device is highly effective despite being more susceptible to noise and using a less direct measurement method. In comparison, the traditional SVM model using wearable data performed lower relative to the more advanced deep learning methods highlighting their limitations when handling complex data.

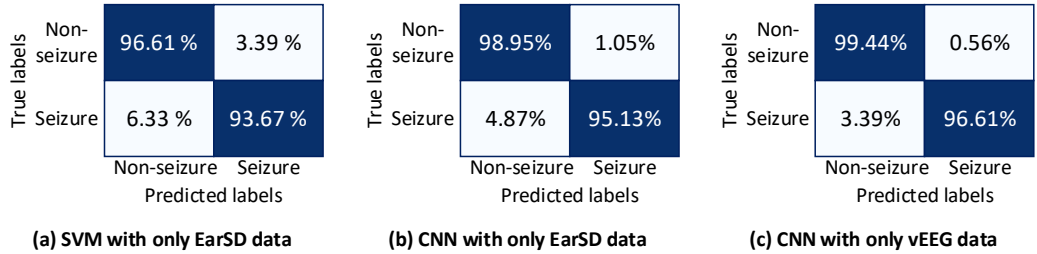

Fig. 20. Comparing the detection performance of the machine learning methods using data from *EarSD* device and the hospital-based vEEG system separately.

While there are challenges in capturing neurological signals from behind the ear, as shown by the slightly reduced accuracy of the *EarSD* device compared to the results using only the vEEG data, we see only a small difference in performance between the systems. This further strengthens our belief that such a device has massive potential to be deployed in the real world and help medical professionals monitor their patients continuously outside of clinical settings for a more comprehensive seizure management system. The goal remains to optimize the wearable technology to match the performance of traditional scalp-based EEG systems, making seizure monitoring more accessible and practical for patients.

Table 4 shows a comparison of *EarSD* against some other seizure detection systems available in the literature. Our comparative study revealed some limitations in existing seizure detection systems. Many are wrist-worn, compromising EEG signal detection due to distance from the source, while head-worn alternatives often prove uncomfortable or impractical for continuous use.

| Systems | Forms | Signal | Algorithm | Sensitivity (%) | #Subjects | Fully-integrated Earable | Low-cost COTS | Motion Artifact Analysis | Patients / Doctors Survey |
|---|---|---|---|---|---|---|---|---|---|
| [120] | Wrist-worn | IMU | RFC | 88.01 | 5 | No | Yes | Low-pass Filter | No |
| [44] | Wrist-worn | ECG, PPG | SVM | 70 | 11 | No | Yes | Filtering (linear, low pass, bandpass) | No |
| [121] | Wrist-worn | EDA, IMU | Gradient Tree Boosting | 91 | 10 | No | Yes | Bandpass Filters | No |
| [122] | Wrist-worn | EDA, IMU, PPG, Temp. | LSTM | 93 | 10 | No | Yes | No | No |
| [43] | Behind ears | EEG, ECG | RFC | 92 | 135 | No | No | No | No |
| [78] | Behind ears | EEG | SVM | 94.5 | 12 | No | No | Bandpass Filters | No |
| [79] | Behind ears | EEG | GAN | 96.3 | 12 | No | No | No | No |
| *EarSD* | Ear-worn | EEG, EMG, EOG | CNN | 95.13 | 32 | Yes | Yes | VMD, NNMF | Yes |

Table 4. *EarSD* and parallel research efforts in wearable-based seizure detection.

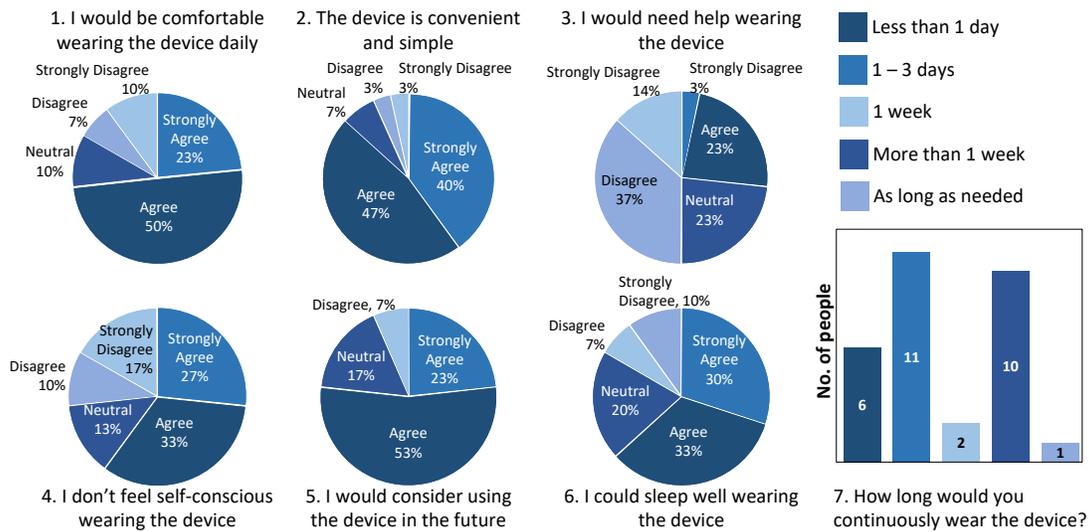

Fig. 21. Responses to User Study Questionnaire showing overwhelmingly positive perception about *EarSD* among the users in our study. This emphasizes the need for an inexpensive, comfortable, and convenient wearable device among patients suffering from epileptic seizures. Devices like *EarSD* can significantly improve the quality of life for such patients.

Additionally, the high cost of commercial systems restricts accessibility for many patients. Addressing these issues, we developed *EarSD*, a lightweight, low-cost, and user-friendly device designed in an earphone form factor using COTS components. This design ensures both affordability and social acceptability, crucial for long-term monitoring as confirmed by positive user feedback in our study (Section 6.3).

*EarSD*'s unique integration of EEG, EMG, and EOG signals provides a comprehensive view of seizure-related physiological changes, distinguishing it from single-modality systems. We employed advanced signal processing techniques like Variational Mode Decomposition (VMD) and Non-Negative Matrix Factorization (NNMF) for enhanced signal quality and reliability. Our clinical trials, conducted with a diverse age range (19-74) of real patients, demonstrate the system's generalizability and robustness. Notably, our Convolutional Neural Network (CNN) approach achieved a high sensitivity of 95.13%, comparable to or surpassing many existing methods. This combination of innovative design, multi-modal signal integration, advanced processing techniques, and strong clinical performance positions our device as a significant advancement in accessible and effective seizure detection technology.

6.3 User Study & Focused Group Discussion

Upon patient discharge, an anonymous, 14-question survey was given to the patients to gather feedback about their experience with *EarSD*. The survey included questions on the users' perspective of *EarSD*'s comfort, ease of use, sleep disruption, social acceptability, and willingness to use such a device. In addition to the anonymous patient survey, a one-hour focus group discussion (FGD) with nine epileptologists was held at the hospital to obtain their opinion on portable seizure detection devices. Conducting such end-user studies allows us to gather insights from experts and patients alike which will help lay the foundations for future improvements. We discuss the results of the survey and the FGD here.

*6.3.1 Survey Results.* Out of the 32 participants who participated in our experiment, 30 (94%) completed the post-survey. Figure 21 presents the questions that we asked our participants to find their opinions and experiences. The results show that 73% found *EarSD* comfortable, while 17% reported discomfort due to pressure and the adhesive glue. An interesting insight that we found was that discomfort was more prevalent among users wearing eyeglasses, likely because both items rely on the ear as a support point. Nevertheless, an overwhelming 87% agreed it was far simpler than conventional hospital EEG setups. A particularly important result was that 80% of the participants indicated that they would be willing to continuously wear the device for prolonged periods (between 3 days to 1 week) and 60% thought the device would be socially acceptable. Finally, 67% of the respondents said it did not hamper their sleep.

In addition to answering the questions in our survey, users provided qualitative feedback on desired features and areas for improvement. Many expressed interest in a user-friendly, self-applicable design to facilitate independent use and enhance comfort and control. Patients were also interested in monitoring their signals through an app, which would provide a summary explanation of their measurements so that they could be more informed about their condition.

*6.3.2 FGD Results.* The purpose of the FGD session was to find out epileptologists' and medical professionals' perspectives on a seizure detection device. Nine epileptologists were invited to a one-hour, recorded focus group to determine the preferences of epileptologists regarding a wearable, EEG-based seizure-detection device. The participants expressed considerable enthusiasm regarding the potential of an EEG-based seizure-detection device like *EarSD*. They acknowledged the immense value of recorded EEG data for their practice, emphasizing that it offers a more dependable source of information compared to patient-reported data. However, the epileptologists did express some reservations and concerns offering valuable suggestions regarding wearable EEG devices. The reliability of the device emerged as a main issue with particular emphasis on the fact that *EarSD* operates as a 2-channel EEG system. Epileptologists stressed the need to maximize the sensitivity and specificity of the device prior to any commercial use as false positives could lead to unnecessary anxiety and possible overmedication for patients.

# 7 LIMITATIONS AND FUTURE DIRECTIONS

## 7.1 Limitations

As a preliminary proof-of-concept study, our results are somewhat limited in terms of our sample size. We had a sample size of 32 patients and larger studies will be needed in the future to better understand the device's generalizability. The strong positive outcome shown by our proposed *EarSD* warrants further investigation through expanded trials to provide a better understanding of the device's generalizability. In addition, while our results have shown promising outputs in detecting seizures, our testing was conducted in controlled hospital settings. It is important to assess real-world factors including external wearables like jewelry and hearing aids, physical conditions such as perspiration, and environmental conditions such as rain, all of which may impact signal quality and electrode contact. Robustness to such potential interferences will be critical for reliable performance outside of clinical settings. Finally, the device was applied using collodion glue to ensure reliable skin contact during this proof-of-concept study, which requires a better solution for long-term usage.

## 7.2 Future Directions

Future iterations of *EarSD* will incorporate the recommendations made by users and epileptologists, especially when it comes to device comfort and ease of use. Further technical improvement is also needed to make the data collection more reliable and feasible in the real world. For example, it is important to have electrodes that can be applied without adhesives like collodion glue and still obtain strong skin-electrode contact to obtain reliable signals. We also plan to improve overall performance and battery life through on-device machine learning, eliminating the need for constant wireless connectivity. A larger, multi-center study will be needed to gather more patient data and minimize sample selection bias. Furthermore, improving the traditional machine learning approaches is difficult since these algorithms depend heavily on feature engineering. The quality of the selected features greatly impacts their performance and identifying the most relevant features for the model can be time-consuming and may require domain expertise. Also, as the dataset increases, training and inference times may become impractically long. Therefore, applying deep learning models could make the system more scaleable and adaptable to complex patterns with their automated feature extraction capabilities. They can also scale much better with larger datasets which can improve detection accuracy and generalization. We also aim to integrate additional capabilities such as classifying seizure types and predicting seizures for proactive interventions. Lastly, the form factor requires continued refinement to improve comfort, aesthetics, and discreetness for easier long-term use. Once optimized, home studies will reveal the strengths and limitations of real-world performance across diverse environments and lifestyles.

## 8 CONCLUSION

In this work, we present *EarSD*, a wearable device designed to enhance the lives of epilepsy patients by providing continuous and at-home monitoring for the detection of seizures. The device contains only two electrodes, worn behind each ear, and records vital physiological signals of EEG, EMG, and EOG, analyzing them to detect seizures and eliminate the need for unnecessary hospital visits. Through our collaboration with a hospital, we were able to test our proposed device on real-world patients and compare it with the *gold-standard* scalp EEG test. Our study involved 32 patients who simultaneously wore the hospital vEEG setup and our *EarSD* device to ensure both devices captured the same events. The recorded signals were preprocessed using our signal processing algorithm to remove noise and extract features. The processed signals were then analyzed using machine learning algorithms of SVM, KNN, RFC, and CNN. We obtained a seizure detection accuracy of 97.9% with CNN using just the recordings of *EarSD* from behind each ear. We also conducted a user study and a focus group discussion with patients and epileptologists to learn the limitations of *EarSD* and receive feedback to further improve the system for future studies. Their responses provide clear directions on the key priorities of end users and lay the foundation for future development. Overall, this work provides substantial evidence that our proposed *EarSD* can reliably capture seizures and contribute to a more effective management of the disease.

## 9 ACKNOWLEDGEMENT

This material is based partly upon work supported by the SONY Faculty Innovation Award (Dr. Phuc Nguyen) and the National Science Foundation under Award Number ECCS #2401415.

## REFERENCES
[1] "Epilepsy," https://www.who.int/news-room/fact-sheets/detail/epilepsy, accessed: 2023-11-02.
[2] R. Barranco, F. Caputo, A. Molinelli, and F. Ventura, "Review on post-mortem diagnosis in suspected SUDEP: Currently still a difficult task for Forensic Pathologists," Journal of Forensic and Legal Medicine, vol. 70, p. 101920, Feb. 2020.
[3] P. F. Prior, R. S. M. Virden, and D. E. Maynard, "An eeg device for monitoring seizure discharges," Epilepsia, vol. 14, 1973. [Online]. Available: https://api.semanticscholar.org/CorpusID:33586341
[4] A. H. Shoeb and J. Guttag, "Application of Machine Learning To Epileptic Seizure Detection," in 2010 International Conference on Machine Learning (ICML), Jun. 2010. [Online]. Available: https://www.semanticscholar.org/paper/Application-of-Machine-Learning-To-Epileptic-Shoeb-Guttag/57e4afe9ca74414fa02f2e0a929b64dc9a03334d
[5] A. S. Zandi, M. Javidan, G. A. Dumont, and R. Tafreshi, "Automated real-time epileptic seizure detection in scalp EEG recordings using an algorithm based on wavelet packet transform," IEEE transactions on bio-medical engineering, vol. 57, no. 7, pp. 1639–1651, Jul. 2010.
[6] B. Blachut, C. Hoppe, R. Surges, C. Elger, and C. Helmstaedter, "Subjective seizure counts by epilepsy clinical drug trial participants are not reliable," Epilepsy & Behavior: E&B, vol. 67, pp. 122–127, Feb. 2017.

[7] "Embrace2 Seizure Monitoring | Smarter Epilepsy Management | Embrace Watch," https://www.empatica.com/embrace2/, accessed: 2023-11-02.

[8] T. Kim, P. Nguyen, N. Pham, N. Bui, H. Truong, S. Ha, and T. Vu, "Epileptic Seizure Detection and Experimental Treatment: A Review," Frontiers in Neurology, vol. 11, 2020. [Online]. Available: https://www.frontiersin.org/articles/10.3389/fneur.2020.00701

[9] BioSerenity, "Neuronaute | bioserenity," https://bioserenity.com/neuronaute/, accessed: 2024-07-22.

[10] Frenz, "Frenz sleep band," https://frenzband.com/, accessed: 2024-07-22.

[11] Emotiv, "Epoc x," https://www.emotiv.com/products/epoc-x, accessed: 2024-07-22.

[12] "EEG - Electroencephalography - BCI | NeuroSky," https://neurosky.com/biosensors/eeg-sensor/, accessed: 2023-11-02.

[13] Macrotellect, "Brainlink," https://o.macrotellect.com/2020/Hardware.html#Pro, accessed: 2024-07-22.

[14] Muse, "Muse," https://choosemuse.com/, accessed: 2024-07-22.

[15] Verus, "Versus mobile eeg," https://getversus.com/, accessed: 2024-07-22.

[16] Neuroon, "Neuroon sleep mask," https://neuroon.com/, accessed: 2024-07-22.

[17] ——, "Naptime - the ultimate power nap assistant," https://www.bestnaptime.com/home, accessed: 2024-07-22.

[18] Z. Liang and T. Nishimura, "Are wearable EEG devices more accurate than fitness wristbands for home sleep Tracking? Comparison of consumer sleep trackers with clinical devices," in 2017 IEEE 6th Global Conference on Consumer Electronics (GCCE), Oct. 2017, pp. 1–5. [Online]. Available: https://ieeexplore.ieee.org/abstract/document/8229188

[19] M. Swash and M. S. Schwartz, Neuromuscular Diseases: A Practical Approach to Diagnosis and Management. Springer Science & Business Media, Mar. 2013, google-Books-ID: ecx9BwAAQBAJ.

[20] K. O'Hearn, M. Asato, S. Ordaz, and B. Luna, "Neurodevelopment and executive function in autism," Development and psychopathology, vol. 20, pp. 1103–32, Feb. 2008.

[21] D. Bathgate, J. S. Snowden, A. Varma, A. Blackshaw, and D. Neary, "Behaviour in frontotemporal dementia, Alzheimer's disease and vascular dementia," Acta Neurologica Scandinavica, vol. 103, no. 6, pp. 367–378, 2001, _eprint: https://onlinelibrary.wiley.com/doi/pdf/10.1034/j.1600-0404.2001.2000236.x. [Online]. Available: https://onlinelibrary.wiley.com/doi/abs/10.1034/j.1600-0404.2001.2000236.x

[22] H. Truong, N. Bui, Z. Raghebi, M. Ceko, N. Pham, P. Nguyen, A. Nguyen, T. Kim, K. Siegfried, E. Stene, T. Tvrdy, L. Weinman, T. Payne, D. Burke, T. Dinh, S. D'Mello, F. Banaei-Kashani, T. Wager, P. Goldstein, and T. Vu, "Painometry: Wearable and objective quantification system for acute postoperative pain," in Proceedings of the 18th International Conference on Mobile Systems, Applications, and Services, ser. MobiSys '20. New York, NY, USA: Association for Computing Machinery, 2020, p. 419–433. [Online]. Available: https://doi.org/10.1145/3386901.3389022

[23] J. Jankovic and E. Tolosa, Parkinson's Disease and Movement Disorders. Lippincott Williams & Wilkins, 2007, google-Books-ID: 0XijYbEuoQcC.

[24] "Video EEG Test," https://www.epilepsy.com/diagnosis/eeg/video-eeg, accessed: 2023-11-02.

[25] T. L. Babb, E. Mariani, and P. H. Crandall, "An electronic circuit for detection of EEG seizures recorded with implanted electrodes," Electroencephalography and Clinical Neurophysiology, vol. 37, no. 3, pp. 305–308, Sep. 1974. [Online]. Available: https://www.sciencedirect.com/science/article/pii/0013469474900364

[26] "Persyst: The worldwide leader in EEG software," https://www.persyst.com/, accessed: 2023-11-02.

[27] M. L. Scheuer, S. B. Wilson, A. Antony, G. Ghearing, A. Urban, and A. I. Bagić, "Seizure Detection: Interreader Agreement and Detection Algorithm Assessments Using a Large Dataset," Journal of Clinical Neurophysiology: Official Publication of the American Electroencephalographic Society, vol. 38, no. 5, pp. 439–447, Sep. 2021.

[28] A. Shoeibi, M. Khodatars, N. Ghassemi, M. Jafari, P. Moridian, R. Alizadehsani, M. Panahiazar, F. Khozeimeh, A. Zare, H. Hosseini-Nejad, A. Khosravi, A. F. Atiya, D. Aminshahidi, S. Hussain, M. Rouhani, S. Nahavandi, and U. R. Acharya, "Epileptic Seizures Detection Using Deep Learning Techniques: A Review," International Journal of Environmental Research and Public Health, vol. 18, no. 11, p. 5780, May 2021. [Online]. Available: https://www.ncbi.nlm.nih.gov/pmc/articles/PMC8199071/

[29] U. Asif, S. Roy, J. Tang, and S. Harrer, "SeizureNet: Multi-Spectral Deep Feature Learning for Seizure Type Classification," Sep. 2020, arXiv:1903.03232 [cs, q-bio, stat]. [Online]. Available: http://arxiv.org/abs/1903.03232

[30] S. Tang, J. A. Dunnmon, K. Saab, X. Zhang, Q. Huang, F. Dubost, D. L. Rubin, and C. Lee-Messer, "Self-Supervised Graph Neural Networks for Improved Electroencephalographic Seizure Analysis," Mar. 2022, arXiv:2104.08336 [cs, eess]. [Online]. Available: http://arxiv.org/abs/2104.08336

[31] S. Roy, U. Asif, J. Tang, and S. Harrer, "Seizure Type Classification Using EEG Signals and Machine Learning: Setting a Benchmark," in 2020 IEEE Signal Processing in Medicine and Biology Symposium (SPMB), Dec. 2020, pp. 1–6, iSSN: 2473-716X. [Online]. Available: https://ieeexplore.ieee.org/document/9353642

[32] I. R. D. Saputro, N. D. Maryati, S. R. Solihati, I. Wijayanto, S. Hadiyoso, and R. Patmasari, "Seizure Type Classification on EEG Signal using Support Vector Machine," Journal of Physics: Conference Series, vol. 1201, no. 1, p. 012065, May 2019, publisher: IOP Publishing. [Online]. Available: https://dx.doi.org/10.1088/1742-6596/1201/1/012065

[33] I. Wijayanto, R. Hartanto, H. A. Nugroho, and B. Winduratna, "Seizure Type Detection in Epileptic EEG Signal using Empirical Mode Decomposition and Support Vector Machine," in 2019 International Seminar on Intelligent Technology and Its Applications (ISITIA), Aug. 2019, pp. 314–319. [Online]. Available: https://ieeexplore.ieee.org/document/8937205

[34] V. Shah, E. von Weltin, S. Lopez, J. R. McHugh, L. Veloso, M. Golmohammadi, I. Obeid, and J. Picone, "The Temple University Hospital Seizure Detection Corpus," Frontiers in Neuroinformatics, vol. 12, 2018. [Online]. Available: https://www.frontiersin.org/articles/10.3389/fninf.2018.00083

[35] S. C. Joshi, G. C. Jana, and A. Agrawal, "A Multi-view Representation Learning Approach for Seizure Detection Over Multi-channel EEG Signals," in Intelligent Data Engineering and Analytics, ser. Smart Innovation, Systems and Technologies, V. Bhateja, X.-S. Yang, J. Chun-Wei Lin, and R. Das, Eds. Singapore: Springer Nature, 2023, pp. 375–385.

[36] A. H. Shoeb, "Application of machine learning to epileptic seizure onset detection and treatment," Thesis, Massachusetts Institute of Technology, 2009, accepted: 2010-04-28T17:17:43Z. [Online]. Available: https://dspace.mit.edu/handle/1721.1/54669

[37] S. Madhavan, R. K. Tripathy, and R. B. Pachori, "Time-Frequency Domain Deep Convolutional Neural Network for the Classification of Focal and Non-Focal EEG Signals," IEEE Sensors Journal, vol. 20, no. 6, pp. 3078–3086, Mar. 2020, conference Name: IEEE Sensors Journal. [Online]. Available: https://ieeexplore.ieee.org/document/8913620

[38] M. Savadkoohi, T. Oladunni, and L. Thompson, "A machine learning approach to epileptic seizure prediction using Electroencephalogram (EEG) Signal," Biocybernetics and Biomedical Engineering, vol. 40, no. 3, pp. 1328–1341, Jul. 2020. [Online]. Available: https://www.sciencedirect.com/science/article/pii/S0208521620300851

[39] L. V. Tran, H. M. Tran, T. M. Le, T. T. M. Huynh, H. T. Tran, and S. V. T. Dao, "Application of Machine Learning in Epileptic Seizure Detection," Diagnostics (Basel, Switzerland), vol. 12, no. 11, p. 2879, Nov. 2022.

[40] D. K. Atal and M. Singh, "Effectual seizure detection using MBBF-GPSO with CNN network," Cognitive Neurodynamics, Feb. 2023. [Online]. Available: https://doi.org/10.1007/s11571-023-09943-1

[41] O. M. Doyle, A. Temko, W. Marnane, G. Lightbody, and G. B. Boylan, "Heart rate based automatic seizure detection in the newborn," Medical Engineering & Physics, vol. 32, no. 8, pp. 829–839, Oct. 2010.

[42] K. Jansen, C. Varon, S. Van Huffel, and L. Lagae, "Peri-ictal ECG changes in childhood epilepsy: implications for detection systems," Epilepsy & Behavior: E&B, vol. 29, no. 1, pp. 72–76, Oct. 2013.

[43] K. Vandecasteele, T. De Cooman, C. Chatzichristos, E. Cleeren, L. Swinnen, J. Macea Ortiz, S. Van Huffel, M. Dümpelmann, A. Schulze-Bonhage, M. De Vos, W. Van Paesschen, and B. Hunyadi, "The power of ECG in multimodal patient-specific seizure monitoring: Added value to an EEG-based detector using limited channels," Epilepsia, vol. 62, no. 10, pp. 2333–2343, Oct. 2021.

[44] K. Vandecasteele, T. De Cooman, Y. Gu, E. Cleeren, K. Claes, W. V. Paesschen, S. V. Huffel, and B. Hunyadi, "Automated Epileptic Seizure Detection Based on Wearable ECG and PPG in a Hospital Environment," Sensors (Basel, Switzerland), vol. 17, no. 10, p. 2338, Oct. 2017.

[45] S. Beniczky, I. Conradsen, and P. Wolf, "Detection of convulsive seizures using surface electromyography," Epilepsia, vol. 59 Suppl 1, pp. 23–29, Jun. 2018.

[46] C. Bagavathi, S. M, S. M. Nair, and S. R, "Novel Epileptic Detection System using Portable EMG-based Assistance," in 2022 International Conference on Applied Artificial Intelligence and Computing (ICAAIC), May 2022, pp. 1762–1765. [Online]. Available: https://ieeexplore.ieee.org/document/9793109

[47] A. Djemal, D. Bouchaala, A. Fakhfakh, and O. Kanoun, "Wearable Electromyography Classification of Epileptic Seizures: A Feasibility Study," Bioengineering, vol. 10, no. 6, p. 703, Jun. 2023, number: 6 Publisher: Multidisciplinary Digital Publishing Institute. [Online]. Available: https://www.mdpi.com/2306-5354/10/6/703

[48] I. C. Zibrandtsen, P. Kidmose, and T. W. Kjaer, "Detection of generalized tonic-clonic seizures from ear-EEG based on EMG analysis," Seizure, vol. 59, pp. 54–59, Jul. 2018. [Online]. Available: https://www.sciencedirect.com/science/article/pii/S1059131118300943

[49] S. Beniczky, I. Conradsen, O. Henning, M. Fabricius, and P. Wolf, "Automated real-time detection of tonic-clonic seizures using a wearable EMG device," Neurology, vol. 90, no. 5, pp. e428–e434, Jan. 2018, publisher: Wolters Kluwer Health, Inc. on behalf of the American Academy of Neurology Section: Article. [Online]. Available: https://n.neurology.org/content/90/5/e428

[50] S. Ganesan, T. A. A. Victoire, and R. Ganesan, "EDA based automatic detection of epileptic seizures using wireless system," in 2011 International Conference on Electronics, Communication and Computing Technologies, Sep. 2011, pp. 47–52. [Online]. Available: https://ieeexplore.ieee.org/document/6077068

[51] M.-Z. Poh, T. Loddenkemper, C. Reinsberger, N. C. Swenson, S. Goyal, M. C. Sabtala, J. R. Madsen, and R. W. Picard, "Convulsive seizure detection using a wrist-worn electrodermal activity and accelerometry biosensor," Epilepsia, vol. 53, no. 5, pp. e93–97, May 2012.

[52] C. A. Szabo', L. C. Morgan, K. M. Karkar, L. D. Leary, O. V. Lie, M. Girouard, and J. E. Cavazos, "Electromyography-based seizure detector: Preliminary results comparing a generalized tonic-clonic seizure detection algorithm to video-EEG recordings," Epilepsia, vol. 56, no. 9, pp. 1432–1437, Sep. 2015.

[53] F. Massé, M. V. Bussel, A. Serteyn, J. Arends, and J. Penders, "Miniaturized wireless ECG monitor for real-time detection of epileptic seizures," ACM Transactions on Embedded Computing Systems, vol. 12, no. 4, pp. 102:1–102:21, Jul. 2013. [Online]. Available: https://dl.acm.org/doi/10.1145/2485984.2485990

[54] J. Jeppesen, A. Fuglsang-Frederiksen, P. Johansen, J. Christensen, S. Wüstenhagen, H. Tankisi, E. Qerama, A. Hess, and S. Beniczky, "Seizure detection based on heart rate variability using a wearable electrocardiography device," Epilepsia, vol. 60, no. 10, pp. 2105–2113, Oct. 2019.

[55] J. Askamp and M. J. A. M. van Putten, "Mobile EEG in epilepsy," International Journal of Psychophysiology, vol. 91, no. 1, pp. 30–35, Jan. 2014. [Online]. Available: https://www.sciencedirect.com/science/article/pii/S0167876013002523

[56] J. Duun-Henriksen, M. Baud, M. P. Richardson, M. Cook, G. Kouvas, J. M. Heasman, D. Friedman, J. Peltola, I. C. Zibrandtsen, and T. W. Kjaer, "A new era in electroencephalographic monitoring? Subscalp devices for ultra-long-term recordings," Epilepsia, vol. 61, no. 9, pp. 1805–1817, Sep. 2020.

[57] B. G. Do Valle, S. S. Cash, and C. G. Sodini, "Wireless behind-the-ear EEG recording device with wireless interface to a mobile device (iPhone/iPod touch)," Annual International Conference of the IEEE Engineering in Medicine and Biology Society. IEEE Engineering in Medicine and Biolo vol. 2014, pp. 5952–5955, 2014.

[58] M. EL Menshawy, A. Benharref, and M. Serhani, "An automatic mobile-health based approach for EEG epileptic seizures detection," Expert Systems with Applications, vol. 42, no. 20, pp. 7157–7174, Nov. 2015. [Online]. Available: https://www.sciencedirect.com/science/article/pii/S0957417415003103

[59] Y. Titgemeyer, R. Surges, D.-M. Altenmüller, S. Fauser, A. Kunze, M. Lanz, M. P. Malter, R. D. Nass, F. von Podewils, J. Remi, S. von Spiczak, A. Strzelczyk, R. M. Ramos, E. Kutafina, and S. M. Jonas, "Can commercially available wearable EEG devices be used for diagnostic purposes? An explorative pilot study," Epilepsy & Behavior, vol. 103, p. 106507, Feb. 2020. [Online]. Available: https://www.sciencedirect.com/science/article/pii/S1525505019306808

[60] A. Biondi, V. Santoro, P. F. Viana, P. Laiou, D. K. Pal, E. Bruno, and M. P. Richardson, "Noninvasive mobile EEG as a tool for seizure monitoring and management: A systematic review," Epilepsia, vol. 63, no. 5, pp. 1041–1063, 2022, _eprint: https://onlinelibrary.wiley.com/doi/pdf/10.1111/epi.17220. [Online]. Available: https://onlinelibrary.wiley.com/doi/abs/10.1111/epi.17220

[61] V. Mihajlović, B. Grundlehner, R. Vullers, and J. Penders, "Wearable, Wireless EEG Solutions in Daily Life Applications: What are we Missing?" IEEE Journal of Biomedical and Health Informatics, vol. 19, no. 1, pp. 6–21, Jan. 2015, conference Name: IEEE Journal of Biomedical and Health Informatics. [Online]. Available: https://ieeexplore.ieee.org/document/6824740

[62] "Remi," https://www.epitel.com/, accessed: 2023-11-02.

[63] "Epihunter," https://www.epihunter.com/, accessed: 2023-11-02.

[64] M. A. Frankel, M. J. Lehmkuhle, M. Watson, K. Fetrow, L. Frey, C. Drees, and M. C. Spitz, "Electrographic seizure monitoring with a novel, wireless, single-channel EEG sensor," Clinical Neurophysiology Practice, vol. 6, pp. 172–178, Jan. 2021. [Online]. Available: https://www.sciencedirect.com/science/article/pii/S2467981X2100024X

[65] "Ceribell Rapid Response EEG," https://ceribell.com/, accessed: 2023-11-02.

[66] F. S. S. Leijten and Dutch TeleEpilepsy Consortium, "Multimodal seizure detection: A review," Epilepsia, vol. 59 Suppl 1, pp. 42–47, Jun. 2018.

[67] K. B. Mikkelsen, S. L. Kappel, D. P. Mandic, and P. Kidmose, "EEG Recorded from the Ear: Characterizing the Ear-EEG Method," Frontiers in Neuroscience, vol. 9, Nov. 2015, publisher: Frontiers. [Online]. Available: https://www.frontiersin.org/journals/neuroscience/articles/10.3389/fnins.2015.00438/full

[68] V. Goverdovsky, D. Looney, P. Kidmose, and D. P. Mandic, "In-Ear EEG From Viscoelastic Generic Earpieces: Robust and Unobtrusive 24/7 Monitoring," IEEE Sensors Journal, vol. 16, no. 1, pp. 271–277, Jan. 2016, conference Name: IEEE Sensors Journal. [Online]. Available: https://ieeexplore.ieee.org/document/7217787/

[69] I. C. Zibrandtsen, P. Kidmose, C. B. Christensen, and T. W. Kjaer, "Ear-EEG detects ictal and interictal abnormalities in focal and generalized epilepsy – A comparison with scalp EEG monitoring," Clinical Neurophysiology, vol. 128, no. 12, pp. 2454–2461, Dec. 2017. [Online]. Available: https://www.sciencedirect.com/science/article/pii/S1388245717310763

[70] R. Kaveh, J. Doong, A. Zhou, C. Schwendeman, K. Gopalan, F. L. Burghardt, A. C. Arias, M. M. Maharbiz, and R. Muller, "Wireless User-Generic Ear EEG," IEEE Transactions on Biomedical Circuits and Systems, vol. 14, no. 4, pp. 727–737, Aug. 2020, conference Name: IEEE Transactions on Biomedical Circuits and Systems. [Online]. Available: https://ieeexplore.ieee.org/document/9115876/?arnumber=9115876

[71] J. Juez, D. Henao, F. Segura, R. Gómez, M. Le Van Quyen, and M. Valderrama, "Development of a wearable system with In-Ear EEG electrodes for the monitoring of brain activities: An application to epilepsy," in 2021 IEEE 2nd International Congress of Biomedical Engineering and Bioengineering (CI-IB&BI), Oct. 2021, pp. 1–4. [Online]. Available: https://ieeexplore.ieee.org/abstract/document/9626123

[72] P. Kidmose, D. Looney, M. Ungstrup, M. L. Rank, and D. P. Mandic, "A Study of Evoked Potentials From Ear-EEG," IEEE Transactions on Biomedical Engineering, vol. 60, no. 10, pp. 2824–2830, Oct. 2013, conference Name: IEEE Transactions on Biomedical Engineering. [Online]. Available: https://ieeexplore.ieee.org/document/6521411

[73] M. G. Bleichner and S. Debener, "Concealed, Unobtrusive Ear-Centered EEG Acquisition: cEEGrids for Transparent EEG," Frontiers in Human Neuroscience, vol. 11, 2017. [Online]. Available: https://www.frontiersin.org/articles/10.3389/fnhum.2017.00163

[74] K. Vandecasteele, T. De Cooman, J. Dan, E. Cleeren, S. Van Huffel, B. Hunyadi, and W. Van Paesschen, "Visual seizure annotation and automated seizure detection using behind-the-ear electroencephalographic channels," Epilepsia, vol. 61, no. 4, pp. 766–775, 2020, _eprint: https://onlinelibrary.wiley.com/doi/pdf/10.1111/epi.16470. [Online]. Available: https://onlinelibrary.wiley.com/doi/abs/10.1111/epi.16470

[75] A. Meiser and M. G. Bleichner, "Ear-EEG compares well to cap-EEG in recording auditory ERPs: a quantification of signal loss," Journal of Neural Engineering, vol. 19, no. 2, p. 026042, Apr. 2022, publisher: IOP Publishing. [Online]. Available: https://dx.doi.org/10.1088/1741-2552/ac5fcb

[76] A. B. Usakli, "Improvement of EEG Signal Acquisition: An Electrical Aspect for State of the Art of Front End," Computational Intelligence and Neuroscience, vol. 2010, p. e630649, Feb. 2010, publisher: Hindawi. [Online]. Available: https://www.hindawi.com/journals/cin/2010/630649/

[77] M. A. Lopez-Gordo, D. Sanchez-Morillo, and F. P. Valle, "Dry EEG Electrodes," Sensors, vol. 14, no. 7, pp. 12 847–12 870, Jul. 2014, number: 7 Publisher: Multidisciplinary Digital Publishing Institute. [Online]. Available: https://www.mdpi.com/1424-8220/14/7/12847

[78] Y. Gu, E. Cleeren, J. Dan, K. Claes, W. Van Paesschen, S. Van Huffel, and B. Hunyadi, "Comparison between Scalp EEG and Behind-the-Ear EEG for Development of a Wearable Seizure Detection System for Patients with Focal Epilepsy," Sensors, vol. 18, no. 1, p. 29, Jan. 2018, number: 1 Publisher: Multidisciplinary Digital Publishing Institute. [Online]. Available: https://www.mdpi.com/1424-8220/18/1/29

[79] S. You, B. H. Cho, S. Yook, J. Y. Kim, Y.-M. Shon, D.-W. Seo, and I. Y. Kim, "Unsupervised automatic seizure detection for focal-onset seizures recorded with behind-the-ear EEG using an anomaly-detecting generative adversarial network," Computer Methods and Programs in Biomedicine, vol. 193, p. 105472, Sep. 2020. [Online]. Available: https://www.sciencedirect.com/science/article/pii/S0169260719320000

[80] J. Ding, Y. Tang, R. Chang, Y. Li, L. Zhang, and F. Yan, "Reduction in the Motion Artifacts in Noncontact ECG Measurements Using a Novel Designed Electrode Structure," Sensors, vol. 23, no. 2, p. 956, Jan. 2023, number: 2 Publisher: Multidisciplinary Digital Publishing Institute. [Online]. Available: https://www.mdpi.com/1424-8220/23/2/956

[81] D. Seok, S. Lee, M. Kim, J. Cho, and C. Kim, "Motion Artifact Removal Techniques for Wearable EEG and PPG Sensor Systems," Frontiers in Electronics, vol. 2, May 2021, publisher: Frontiers. [Online]. Available: https://www.frontiersin.org/journals/electronics/articles/10.3389/felec.2021.685513/full

[82] Kaya, "A Brief Summary of EEG Artifact Handling," in Brain-Computer Interface. IntechOpen, Jul. 2021. [Online]. Available: https://www.intechopen.com/chapters/77731

[83] G. S. Spencer, J. A. Smith, M. E. H. Chowdhury, R. Bowtell, and K. J. Mullinger, "Exploring the origins of EEG motion artefacts during simultaneous fMRI acquisition: Implications for motion artefact correction," NeuroImage, vol. 173, pp. 188–198, Jun. 2018. [Online]. Available: https://www.sciencedirect.com/science/article/pii/S1053811918301319

[84] J. J. Falco-Walter, I. E. Scheffer, and R. S. Fisher, "The new definition and classification of seizures and epilepsy," Epilepsy Research, vol. 139, pp. 73–79, Jan. 2018. [Online]. Available: https://www.sciencedirect.com/science/article/pii/S0920121117303819

[85] E. Tuncer and E. D. Bolat, "Channel based epilepsy seizure type detection from electroencephalography (EEG) signals with machine learning techniques," Biocybernetics and Biomedical Engineering, vol. 42, no. 2, pp. 575–595, Apr. 2022. [Online]. Available: https://www.sciencedirect.com/science/article/pii/S0208521622000365

[86] H. de Talhouet and J. G. Webster, "The origin of skin-stretch-caused motion artifacts under electrodes," Physiological Measurement, vol. 17, no. 2, pp. 81–93, May 1996.

[87] A. Casson, "Gelatine characterization for electrophysiology phantoms," Feb. 2021. [Online]. Available: https://data.mendeley.com/datasets/74z6cc3j63/2

[88] J.-T. Chien, "Chapter 4 - Independent Component Analysis," in Source Separation and Machine Learning, J.-T. Chien, Ed. Academic Press, Jan. 2019, pp. 99–160. [Online]. Available: https://www.sciencedirect.com/science/article/pii/B9780128045664000164

[89] J. Xu, S. Mitra, C. Van Hoof, R. F. Yazicioglu, and K. A. A. Makinwa, "Active Electrodes for Wearable EEG Acquisition: Review and Electronics Design Methodology," IEEE reviews in biomedical engineering, vol. 10, pp. 187–198, 2017.

[90] N. E. Huang, Z. Shen, S. R. Long, M. C. Wu, H. H. Shih, Q. Zheng, N.-C. Yen, C. C. Tung, and H. H. Liu, "The empirical mode decomposition and the Hilbert spectrum for nonlinear and non-stationary time series analysis," Proceedings of the Royal Society of London. Series A: Mathematical, Physical and Engineering Sciences, vol. 454, no. 1971, pp. 903–995, Mar. 1998, publisher: Royal Society. [Online]. Available: https://royalsocietypublishing.org/doi/10.1098/rspa.1998.0193

[91] J. Chen, Y. Huang, and J. Benesty, "Filtering Techniques for Noise Reduction and Speech Enhancement," in Adaptive Signal Processing: Applications to Real-World Problems, ser. Signals and Communication Technology, J. Benesty and Y. Huang, Eds. Berlin, Heidelberg: Springer, 2003, pp. 129–154. [Online]. Available: https://doi.org/10.1007/978-3-662-11028-7_5

[92] D. Butusov, T. Karimov, A. Voznesenskiy, D. Kaplun, V. Andreev, and V. Ostrovskii, "Filtering Techniques for Chaotic Signal Processing," Electronics, vol. 7, no. 12, p. 450, Dec. 2018, number: 12 Publisher: Multidisciplinary Digital Publishing Institute. [Online]. Available: https://www.mdpi.com/2079-9292/7/12/450

[93] J. V. Stone, "Independent component analysis: an introduction," Trends in Cognitive Sciences, vol. 6, no. 2, pp. 59–64, Feb. 2002. [Online]. Available: https://www.sciencedirect.com/science/article/pii/S1364661300018131

[94] G. Naik and D. Kumar, "An Overview of Independent Component Analysis and Its Applications," Informatica, vol. 35, pp. 63–81, Jan. 2011.

[95] H. Kasban, H. Arafa, and S. M. S. Elaraby, "Principle component analysis for radiotracer signal separation," Applied Radiation and Isotopes, vol. 112, pp. 20–26, Jun. 2016. [Online]. Available: https://www.sciencedirect.com/science/article/pii/S0969804316300896

[96] M. A. Kass and Y. Li, "Use of principal component analysis in the de-noising and signal- separation of transient electromagnetic data," 2007. [Online]. Available: https://api.semanticscholar.org/CorpusID:17798116

[97] G. Rilling, P. Flandrin, and P. Gonçalves, "On empirical mode decomposition and its algorithms," Jun. 2003. [Online]. Available: https://www.semanticscholar.org/paper/On-empirical-mode-decomposition-and-its-algorithms-Rilling-Flandrin/3f616db40f5da4446a039bb6ae5d801d4c616f2b

[98] G. Wang, X.-Y. Chen, F.-L. Qiao, Z. Wu, and N. E. Huang, "On intrinsic mode function," Advances in Adaptive Data Analysis, vol. 02, no. 03, pp. 277–293, Jul. 2010, publisher: World Scientific Publishing Co. [Online]. Available: https://www.worldscientific.com/doi/abs/10.1142/S1793536910000549

[99] D. D. Lee and H. S. Seung, "Learning the parts of objects by non-negative matrix factorization," Nature, vol. 401, no. 6755, pp. 788–791, Oct. 1999, number: 6755 Publisher: Nature Publishing Group. [Online]. Available: https://www.nature.com/articles/44565

[100] F. Segovia, J. M. Górriz, J. Ramírez, F. J. Martinez-Murcia, and M. García-Pérez, "Using deep neural networks along with dimensionality reduction techniques to assist the diagnosis of neurodegenerative disorders," Logic Journal of the IGPL, vol. 26, no. 6, pp. 618–628, Nov. 2018. [Online]. Available: https://doi.org/10.1093/jigpal/jzy026

[101] S. Krause-Solberg, "Non-Negative Dimensionality Reduction in Signal Separation," doctoralThesis, Staats- und Universitätsbibliothek Hamburg Carl von Ossietzky, 2015, accepted: 2020-10-19T13:15:27Z Journal Abbreviation: Anwendung von nichtnegativer Dimensionsreduktion im Bereich der Signaltrennung. [Online]. Available: https://ediss.sub.uni-hamburg.de/handle/ediss/6859

[102] B. Karan, S. S. Sahu, J. R. Orozco-Arroyave, and K. Mahto, "Non-negative matrix factorization-based time-frequency feature extraction of voice signal for Parkinson's disease prediction," Computer Speech & Language, vol. 69, p. 101216, Sep. 2021. [Online]. Available: https://www.sciencedirect.com/science/article/pii/S0885230821000231

[103] Y. Yi, Y. Shi, H. Zhang, J. Wang, and J. Kong, "Label propagation based semi-supervised non-negative matrix factorization for feature extraction," Neurocomputing, vol. 149, pp. 1021–1037, Feb. 2015. [Online]. Available: https://www.sciencedirect.com/science/article/pii/S0925231214009680

[104] E. Vincent, R. Gribonval, and C. Fevotte, "Performance measurement in blind audio source separation," IEEE Transactions on Audio, Speech, and Language Processing, vol. 14, no. 4, pp. 1462–1469, Jul. 2006, conference Name: IEEE Transactions on Audio, Speech, and Language Processing. [Online]. Available: https://ieeexplore.ieee.org/document/1643671

[105] C. Ye, K. Toyoda, and T. Ohtsuki, "Blind Source Separation on Non-Contact Heartbeat Detection by Non-Negative Matrix Factorization Algorithms," IEEE Transactions on Biomedical Engineering, vol. 67, no. 2, pp. 482–494, Feb. 2020, conference Name: IEEE Transactions on Biomedical Engineering. [Online]. Available: https://ieeexplore.ieee.org/abstract/document/8710248

[106] A. Cichocki, R. Zdunek, and S. Amari, "New Algorithms for Non-Negative Matrix Factorization in Applications to Blind Source Separation," in 2006 IEEE International Conference on Acoustics Speech and Signal Processing Proceedings, vol. 5, May 2006, pp. V–V, iSSN: 2379-190X. [Online]. Available: https://ieeexplore.ieee.org/document/1661352

[107] B. S. Alexandrov and V. V. Vesselinov, "Blind source separation for groundwater pressure analysis based on nonnegative matrix factorization," Water Resources Research, vol. 50, no. 9, pp. 7332–7347, 2014, _eprint: https://onlinelibrary.wiley.com/doi/pdf/10.1002/2013WR015037. [Online]. Available: https://onlinelibrary.wiley.com/doi/abs/10.1002/2013WR015037

[108] A. Rolet, V. Seguy, M. Blondel, and H. Sawada, "Blind source separation with optimal transport non-negative matrix factorization," EURASIP Journal on Advances in Signal Processing, vol. 2018, no. 1, p. 53, Sep. 2018. [Online]. Available: https://doi.org/10.1186/s13634-018-0576-2

[109] L. Jing, C. Zhang, and M. K. Ng, "SNMFCA: Supervised NMF-Based Image Classification and Annotation," IEEE Transactions on Image Processing, vol. 21, no. 11, pp. 4508–4521, Nov. 2012, conference Name: IEEE Transactions on Image Processing. [Online]. Available: https://ieeexplore.ieee.org/document/6226461

[110] C. Févotte, N. Bertin, and J.-L. Durrieu, "Nonnegative Matrix Factorization with the Itakura-Saito Divergence: With Application to Music Analysis," Neural Computation, vol. 21, no. 3, pp. 793–830, Mar. 2009. [Online]. Available: https://doi.org/10.1162/neco.2008.04-08-771

[111] V. Leplat, N. Gillis, and J. Idier, "Multiplicative Updates for NMF with $\beta$-Divergences under Disjoint Equality Constraints," SIAM Journal on Matrix Analysis and Applications, vol. 42, no. 2, pp. 730–752, Jan. 2021, publisher: Society for Industrial and Applied Mathematics. [Online]. Available: https://epubs.siam.org/doi/abs/10.1137/20M1377278

[112] C. Févotte and J. Idier, "Algorithms for Nonnegative Matrix Factorization with the Beta-Divergence," Neural Computation, vol. 23, no. 9, pp. 2421–2456, Sep. 2011. [Online]. Available: https://doi.org/10.1162/NECO_a_00168

[113] N. N. E. Software, "Natus® NeuroWorks® EEG Software," Nov. 2023. [Online]. Available: https://natus.com/neuro/neuroworks-eeg-software/

[114] P. L. Nunez and R. Srinivasan, Electric Fields of the Brain.  Oxford University Press, Jan. 2006. [Online]. Available: https://academic.oup.com/book/2998

[115] L. J. Hirsch, M. W. Fong, M. Leitinger, S. M. LaRoche, S. Beniczky, N. S. Abend, J. W. Lee, C. J. Wusthoff, C. D. Hahn, M. B. Westover, E. E. Gerard, S. T. Herman, H. A. Haider, G. Osman, A. Rodriguez-Ruiz, C. B. Maciel, E. J. Gilmore, A. Fernandez, E. S. Rosenthal, J. Claassen, A. M. Husain, J. Y. Yoo, E. L. So, P. W. Kaplan, M. R. Nuwer, M. van Putten, R. Sutter, F. W. Drislane, E. Trinka, and N. Gaspard, "American Clinical Neurophysiology Society's Standardized Critical Care EEG Terminology: 2021 Version," Journal of clinical neurophysiology : official publication of the American Electroencephalographic Society, vol. 38, no. 1, pp. 1–29, Jan. 2021. [Online]. Available: https://www.ncbi.nlm.nih.gov/pmc/articles/PMC8135051/

[116] X. Zhao, N. Yoshida, T. Ueda, H. Sugano, and T. Tanaka, "Epileptic seizure detection by using interpretable machine learning models," Journal of Neural Engineering, vol. 20, no. 1, p. 015002, Feb. 2023, publisher: IOP Publishing. [Online]. Available: https://dx.doi.org/10.1088/1741-2552/acb089

[117] Z. Sawadogo, G. Mendy, J. M. Dembele, and S. Ouya, "Android malware detection: Investigating the impact of imbalanced data-sets on the performance of machine learning models," in 2022 24th International Conference on Advanced Communication Technology (ICACT), Feb. 2022, pp. 435–441, iSSN: 1738-9445. [Online]. Available: https://ieeexplore.ieee.org/document/9728833

[118] O. Medical, "Visensia," 2023. [Online]. Available: https://www.obsmedical.com/visensia-the-safety-index/

[119] B. Company Inc., "BrainScope," Nov. 2023. [Online]. Available: https://www.brainscope.com

[120] C. Dong, L. Chen, T. Ye, X. Long, R. M. Aarts, J. van Dijk, C. Shang, X. Liao, and Y. Wang, "Home-based Detection of Epileptic Seizures Using a Bracelet with Motor Sensors," in 2021 10th International IEEE/EMBS Conference on Neural Engineering (NER), May 2021, pp. 854–857, iSSN: 1948-3554. [Online]. Available: https://ieeexplore.ieee.org/document/9441171/?arnumber=9441171

[121] S. Böttcher, E. Bruno, N. V. Manyakov, N. Epitashvili, K. Claes, M. Glasstetter, S. Thorpe, S. Lees, M. Dümpelmann, K. V. Laerhoven, M. P. Richardson, A. Schulze-Bonhage, and T. R.-C. Consortium, "Detecting Tonic-Clonic Seizures in Multimodal Biosignal Data From Wearables: Methodology Design and Validation," JMIR mHealth and uHealth, vol. 9, no. 11, p. e27674, Nov. 2021, company: JMIR mHealth and uHealth Distributor: JMIR mHealth and uHealth Institution: JMIR mHealth and uHealth Label: JMIR mHealth and uHealth Publisher: JMIR Publications Inc., Toronto, Canada. [Online]. Available: https://mhealth.jmir.org/2021/11/e27674

[122] M. Nasseri, T. Pal Attia, B. Joseph, N. M. Gregg, E. S. Nurse, P. F. Viana, A. Schulze-Bonhage, M. Dümpelmann, G. Worrell, D. R. Freestone, M. P. Richardson, and B. H. Brinkmann, "Non-invasive wearable seizure detection using long–short-term memory networks with transfer learning," Journal of Neural Engineering, vol. 18, no. 5, p. 056017, Oct. 2021. [Online]. Available: https://iopscience.iop.org/article/10.1088/1741-2552/abef8a